\documentclass[useAMS,usenatbib]{mn2e}

\usepackage{graphicx}
\usepackage{float}
\usepackage{times}
\usepackage{amsfonts,amssymb}
\usepackage{amsmath}
\usepackage{color}
\definecolor{AliceBlue}{rgb}{0.94,0.97,1.00}
\definecolor{AntiqueWhite1}{rgb}{1.00,0.94,0.86}
\definecolor{AntiqueWhite2}{rgb}{0.93,0.87,0.80}
\definecolor{AntiqueWhite3}{rgb}{0.80,0.75,0.69}
\definecolor{AntiqueWhite4}{rgb}{0.55,0.51,0.47}
\definecolor{AntiqueWhite}{rgb}{0.98,0.92,0.84}
\definecolor{BlanchedAlmond}{rgb}{1.00,0.92,0.80}
\definecolor{BlueViolet}{rgb}{0.54,0.17,0.89}
\definecolor{CadetBlue1}{rgb}{0.60,0.96,1.00}
\definecolor{CadetBlue2}{rgb}{0.56,0.90,0.93}
\definecolor{CadetBlue3}{rgb}{0.48,0.77,0.80}
\definecolor{CadetBlue4}{rgb}{0.33,0.53,0.55}
\definecolor{CadetBlue}{rgb}{0.37,0.62,0.63}
\definecolor{CornflowerBlue}{rgb}{0.39,0.58,0.93}
\definecolor{DarkBlue}{rgb}{0.00,0.00,0.55}
\definecolor{DarkCyan}{rgb}{0.00,0.55,0.55}
\definecolor{DarkGoldenrod1}{rgb}{1.00,0.73,0.06}
\definecolor{DarkGoldenrod2}{rgb}{0.93,0.68,0.05}
\definecolor{DarkGoldenrod3}{rgb}{0.80,0.58,0.05}
\definecolor{DarkGoldenrod4}{rgb}{0.55,0.40,0.03}
\definecolor{DarkGoldenrod}{rgb}{0.72,0.53,0.04}
\definecolor{DarkGray}{rgb}{0.66,0.66,0.66}
\definecolor{DarkGreen}{rgb}{0.00,0.39,0.00}
\definecolor{DarkGrey}{rgb}{0.66,0.66,0.66}
\definecolor{DarkKhaki}{rgb}{0.74,0.72,0.42}
\definecolor{DarkMagenta}{rgb}{0.55,0.00,0.55}
\definecolor{DarkOliveGreen1}{rgb}{0.79,1.00,0.44}
\definecolor{DarkOliveGreen2}{rgb}{0.74,0.93,0.41}
\definecolor{DarkOliveGreen3}{rgb}{0.64,0.80,0.35}
\definecolor{DarkOliveGreen4}{rgb}{0.43,0.55,0.24}
\definecolor{DarkOliveGreen}{rgb}{0.33,0.42,0.18}
\definecolor{DarkOrange1}{rgb}{1.00,0.50,0.00}
\definecolor{DarkOrange2}{rgb}{0.93,0.46,0.00}
\definecolor{DarkOrange3}{rgb}{0.80,0.40,0.00}
\definecolor{DarkOrange4}{rgb}{0.55,0.27,0.00}
\definecolor{DarkOrange}{rgb}{1.00,0.55,0.00}
\definecolor{DarkOrchid1}{rgb}{0.75,0.24,1.00}
\definecolor{DarkOrchid2}{rgb}{0.70,0.23,0.93}
\definecolor{DarkOrchid3}{rgb}{0.60,0.20,0.80}
\definecolor{DarkOrchid4}{rgb}{0.41,0.13,0.55}
\definecolor{DarkOrchid}{rgb}{0.60,0.20,0.80}
\definecolor{DarkRed}{rgb}{0.55,0.00,0.00}
\definecolor{DarkSalmon}{rgb}{0.91,0.59,0.48}
\definecolor{DarkSeaGreen1}{rgb}{0.76,1.00,0.76}
\definecolor{DarkSeaGreen2}{rgb}{0.71,0.93,0.71}
\definecolor{DarkSeaGreen3}{rgb}{0.61,0.80,0.61}
\definecolor{DarkSeaGreen4}{rgb}{0.41,0.55,0.41}
\definecolor{DarkSeaGreen}{rgb}{0.56,0.74,0.56}
\definecolor{DarkSlateBlue}{rgb}{0.28,0.24,0.55}
\definecolor{DarkSlateGray1}{rgb}{0.59,1.00,1.00}
\definecolor{DarkSlateGray2}{rgb}{0.55,0.93,0.93}
\definecolor{DarkSlateGray3}{rgb}{0.47,0.80,0.80}
\definecolor{DarkSlateGray4}{rgb}{0.32,0.55,0.55}
\definecolor{DarkSlateGray}{rgb}{0.18,0.31,0.31}
\definecolor{DarkSlateGrey}{rgb}{0.18,0.31,0.31}
\definecolor{DarkTurquoise}{rgb}{0.00,0.81,0.82}
\definecolor{DarkViolet}{rgb}{0.58,0.00,0.83}
\definecolor{DeepPink1}{rgb}{1.00,0.08,0.58}
\definecolor{DeepPink2}{rgb}{0.93,0.07,0.54}
\definecolor{DeepPink3}{rgb}{0.80,0.06,0.46}
\definecolor{DeepPink4}{rgb}{0.55,0.04,0.31}
\definecolor{DeepPink}{rgb}{1.00,0.08,0.58}
\definecolor{DeepSkyBlue1}{rgb}{0.00,0.75,1.00}
\definecolor{DeepSkyBlue2}{rgb}{0.00,0.70,0.93}
\definecolor{DeepSkyBlue3}{rgb}{0.00,0.60,0.80}
\definecolor{DeepSkyBlue4}{rgb}{0.00,0.41,0.55}
\definecolor{DeepSkyBlue}{rgb}{0.00,0.75,1.00}
\definecolor{DimGray}{rgb}{0.41,0.41,0.41}
\definecolor{DimGrey}{rgb}{0.41,0.41,0.41}
\definecolor{DodgerBlue1}{rgb}{0.12,0.56,1.00}
\definecolor{DodgerBlue2}{rgb}{0.11,0.53,0.93}
\definecolor{DodgerBlue3}{rgb}{0.09,0.45,0.80}
\definecolor{DodgerBlue4}{rgb}{0.06,0.31,0.55}
\definecolor{DodgerBlue}{rgb}{0.12,0.56,1.00}
\definecolor{FloralWhite}{rgb}{1.00,0.98,0.94}
\definecolor{ForestGreen}{rgb}{0.13,0.55,0.13}
\definecolor{GhostWhite}{rgb}{0.97,0.97,1.00}
\definecolor{GreenYellow}{rgb}{0.68,1.00,0.18}
\definecolor{HotPink1}{rgb}{1.00,0.43,0.71}
\definecolor{HotPink2}{rgb}{0.93,0.42,0.65}
\definecolor{HotPink3}{rgb}{0.80,0.38,0.56}
\definecolor{HotPink4}{rgb}{0.55,0.23,0.38}
\definecolor{HotPink}{rgb}{1.00,0.41,0.71}
\definecolor{IndianRed1}{rgb}{1.00,0.42,0.42}
\definecolor{IndianRed2}{rgb}{0.93,0.39,0.39}
\definecolor{IndianRed3}{rgb}{0.80,0.33,0.33}
\definecolor{IndianRed4}{rgb}{0.55,0.23,0.23}
\definecolor{IndianRed}{rgb}{0.80,0.36,0.36}
\definecolor{LavenderBlush1}{rgb}{1.00,0.94,0.96}
\definecolor{LavenderBlush2}{rgb}{0.93,0.88,0.90}
\definecolor{LavenderBlush3}{rgb}{0.80,0.76,0.77}
\definecolor{LavenderBlush4}{rgb}{0.55,0.51,0.53}
\definecolor{LavenderBlush}{rgb}{1.00,0.94,0.96}
\definecolor{LawnGreen}{rgb}{0.49,0.99,0.00}
\definecolor{LemonChiffon1}{rgb}{1.00,0.98,0.80}
\definecolor{LemonChiffon2}{rgb}{0.93,0.91,0.75}
\definecolor{LemonChiffon3}{rgb}{0.80,0.79,0.65}
\definecolor{LemonChiffon4}{rgb}{0.55,0.54,0.44}
\definecolor{LemonChiffon}{rgb}{1.00,0.98,0.80}
\definecolor{LightBlue1}{rgb}{0.75,0.94,1.00}
\definecolor{LightBlue2}{rgb}{0.70,0.87,0.93}
\definecolor{LightBlue3}{rgb}{0.60,0.75,0.80}
\definecolor{LightBlue4}{rgb}{0.41,0.51,0.55}
\definecolor{LightBlue}{rgb}{0.68,0.85,0.90}
\definecolor{LightCoral}{rgb}{0.94,0.50,0.50}
\definecolor{LightCyan1}{rgb}{0.88,1.00,1.00}
\definecolor{LightCyan2}{rgb}{0.82,0.93,0.93}
\definecolor{LightCyan3}{rgb}{0.71,0.80,0.80}
\definecolor{LightCyan4}{rgb}{0.48,0.55,0.55}
\definecolor{LightCyan}{rgb}{0.88,1.00,1.00}
\definecolor{LightGoldenrod1}{rgb}{1.00,0.93,0.55}
\definecolor{LightGoldenrod2}{rgb}{0.93,0.86,0.51}
\definecolor{LightGoldenrod3}{rgb}{0.80,0.75,0.44}
\definecolor{LightGoldenrod4}{rgb}{0.55,0.51,0.30}
\definecolor{LightGoldenrodYellow}{rgb}{0.98,0.98,0.82}
\definecolor{LightGoldenrod}{rgb}{0.93,0.87,0.51}
\definecolor{LightGray}{rgb}{0.83,0.83,0.83}
\definecolor{LightGreen}{rgb}{0.56,0.93,0.56}
\definecolor{LightGrey}{rgb}{0.83,0.83,0.83}
\definecolor{LightPink1}{rgb}{1.00,0.68,0.73}
\definecolor{LightPink2}{rgb}{0.93,0.64,0.68}
\definecolor{LightPink3}{rgb}{0.80,0.55,0.58}
\definecolor{LightPink4}{rgb}{0.55,0.37,0.40}
\definecolor{LightPink}{rgb}{1.00,0.71,0.76}
\definecolor{LightSalmon1}{rgb}{1.00,0.63,0.48}
\definecolor{LightSalmon2}{rgb}{0.93,0.58,0.45}
\definecolor{LightSalmon3}{rgb}{0.80,0.51,0.38}
\definecolor{LightSalmon4}{rgb}{0.55,0.34,0.26}
\definecolor{LightSalmon}{rgb}{1.00,0.63,0.48}
\definecolor{LightSeaGreen}{rgb}{0.13,0.70,0.67}
\definecolor{LightSkyBlue1}{rgb}{0.69,0.89,1.00}
\definecolor{LightSkyBlue2}{rgb}{0.64,0.83,0.93}
\definecolor{LightSkyBlue3}{rgb}{0.55,0.71,0.80}
\definecolor{LightSkyBlue4}{rgb}{0.38,0.48,0.55}
\definecolor{LightSkyBlue}{rgb}{0.53,0.81,0.98}
\definecolor{LightSlateBlue}{rgb}{0.52,0.44,1.00}
\definecolor{LightSlateGray}{rgb}{0.47,0.53,0.60}
\definecolor{LightSlateGrey}{rgb}{0.47,0.53,0.60}
\definecolor{LightSteelBlue1}{rgb}{0.79,0.88,1.00}
\definecolor{LightSteelBlue2}{rgb}{0.74,0.82,0.93}
\definecolor{LightSteelBlue3}{rgb}{0.64,0.71,0.80}
\definecolor{LightSteelBlue4}{rgb}{0.43,0.48,0.55}
\definecolor{LightSteelBlue}{rgb}{0.69,0.77,0.87}
\definecolor{LightYellow1}{rgb}{1.00,1.00,0.88}
\definecolor{LightYellow2}{rgb}{0.93,0.93,0.82}
\definecolor{LightYellow3}{rgb}{0.80,0.80,0.71}
\definecolor{LightYellow4}{rgb}{0.55,0.55,0.48}
\definecolor{LightYellow}{rgb}{1.00,1.00,0.88}
\definecolor{LimeGreen}{rgb}{0.20,0.80,0.20}
\definecolor{MediumAquamarine}{rgb}{0.40,0.80,0.67}
\definecolor{MediumBlue}{rgb}{0.00,0.00,0.80}
\definecolor{MediumOrchid1}{rgb}{0.88,0.40,1.00}
\definecolor{MediumOrchid2}{rgb}{0.82,0.37,0.93}
\definecolor{MediumOrchid3}{rgb}{0.71,0.32,0.80}
\definecolor{MediumOrchid4}{rgb}{0.48,0.22,0.55}
\definecolor{MediumOrchid}{rgb}{0.73,0.33,0.83}
\definecolor{MediumPurple1}{rgb}{0.67,0.51,1.00}
\definecolor{MediumPurple2}{rgb}{0.62,0.47,0.93}
\definecolor{MediumPurple3}{rgb}{0.54,0.41,0.80}
\definecolor{MediumPurple4}{rgb}{0.36,0.28,0.55}
\definecolor{MediumPurple}{rgb}{0.58,0.44,0.86}
\definecolor{MediumSeaGreen}{rgb}{0.24,0.70,0.44}
\definecolor{MediumSlateBlue}{rgb}{0.48,0.41,0.93}
\definecolor{MediumSpringGreen}{rgb}{0.00,0.98,0.60}
\definecolor{MediumTurquoise}{rgb}{0.28,0.82,0.80}
\definecolor{MediumVioletRed}{rgb}{0.78,0.08,0.52}
\definecolor{MidnightBlue}{rgb}{0.10,0.10,0.44}
\definecolor{MintCream}{rgb}{0.96,1.00,0.98}
\definecolor{MistyRose1}{rgb}{1.00,0.89,0.88}
\definecolor{MistyRose2}{rgb}{0.93,0.84,0.82}
\definecolor{MistyRose3}{rgb}{0.80,0.72,0.71}
\definecolor{MistyRose4}{rgb}{0.55,0.49,0.48}
\definecolor{MistyRose}{rgb}{1.00,0.89,0.88}
\definecolor{NavajoWhite1}{rgb}{1.00,0.87,0.68}
\definecolor{NavajoWhite2}{rgb}{0.93,0.81,0.63}
\definecolor{NavajoWhite3}{rgb}{0.80,0.70,0.55}
\definecolor{NavajoWhite4}{rgb}{0.55,0.47,0.37}
\definecolor{NavajoWhite}{rgb}{1.00,0.87,0.68}
\definecolor{NavyBlue}{rgb}{0.00,0.00,0.50}
\definecolor{OldLace}{rgb}{0.99,0.96,0.90}
\definecolor{OliveDrab1}{rgb}{0.75,1.00,0.24}
\definecolor{OliveDrab2}{rgb}{0.70,0.93,0.23}
\definecolor{OliveDrab3}{rgb}{0.60,0.80,0.20}
\definecolor{OliveDrab4}{rgb}{0.41,0.55,0.13}
\definecolor{OliveDrab}{rgb}{0.42,0.56,0.14}
\definecolor{OrangeRed1}{rgb}{1.00,0.27,0.00}
\definecolor{OrangeRed2}{rgb}{0.93,0.25,0.00}
\definecolor{OrangeRed3}{rgb}{0.80,0.22,0.00}
\definecolor{OrangeRed4}{rgb}{0.55,0.15,0.00}
\definecolor{OrangeRed}{rgb}{1.00,0.27,0.00}
\definecolor{PaleGoldenrod}{rgb}{0.93,0.91,0.67}
\definecolor{PaleGreen1}{rgb}{0.60,1.00,0.60}
\definecolor{PaleGreen2}{rgb}{0.56,0.93,0.56}
\definecolor{PaleGreen3}{rgb}{0.49,0.80,0.49}
\definecolor{PaleGreen4}{rgb}{0.33,0.55,0.33}
\definecolor{PaleGreen}{rgb}{0.60,0.98,0.60}
\definecolor{PaleTurquoise1}{rgb}{0.73,1.00,1.00}
\definecolor{PaleTurquoise2}{rgb}{0.68,0.93,0.93}
\definecolor{PaleTurquoise3}{rgb}{0.59,0.80,0.80}
\definecolor{PaleTurquoise4}{rgb}{0.40,0.55,0.55}
\definecolor{PaleTurquoise}{rgb}{0.69,0.93,0.93}
\definecolor{PaleVioletRed1}{rgb}{1.00,0.51,0.67}
\definecolor{PaleVioletRed2}{rgb}{0.93,0.47,0.62}
\definecolor{PaleVioletRed3}{rgb}{0.80,0.41,0.54}
\definecolor{PaleVioletRed4}{rgb}{0.55,0.28,0.36}
\definecolor{PaleVioletRed}{rgb}{0.86,0.44,0.58}
\definecolor{PapayaWhip}{rgb}{1.00,0.94,0.84}
\definecolor{PeachPuff1}{rgb}{1.00,0.85,0.73}
\definecolor{PeachPuff2}{rgb}{0.93,0.80,0.68}
\definecolor{PeachPuff3}{rgb}{0.80,0.69,0.58}
\definecolor{PeachPuff4}{rgb}{0.55,0.47,0.40}
\definecolor{PeachPuff}{rgb}{1.00,0.85,0.73}
\definecolor{PowderBlue}{rgb}{0.69,0.88,0.90}
\definecolor{RosyBrown1}{rgb}{1.00,0.76,0.76}
\definecolor{RosyBrown2}{rgb}{0.93,0.71,0.71}
\definecolor{RosyBrown3}{rgb}{0.80,0.61,0.61}
\definecolor{RosyBrown4}{rgb}{0.55,0.41,0.41}
\definecolor{RosyBrown}{rgb}{0.74,0.56,0.56}
\definecolor{RoyalBlue1}{rgb}{0.28,0.46,1.00}
\definecolor{RoyalBlue2}{rgb}{0.26,0.43,0.93}
\definecolor{RoyalBlue3}{rgb}{0.23,0.37,0.80}
\definecolor{RoyalBlue4}{rgb}{0.15,0.25,0.55}
\definecolor{RoyalBlue}{rgb}{0.25,0.41,0.88}
\definecolor{SaddleBrown}{rgb}{0.55,0.27,0.07}
\definecolor{SandyBrown}{rgb}{0.96,0.64,0.38}
\definecolor{SeaGreen1}{rgb}{0.33,1.00,0.62}
\definecolor{SeaGreen2}{rgb}{0.31,0.93,0.58}
\definecolor{SeaGreen3}{rgb}{0.26,0.80,0.50}
\definecolor{SeaGreen4}{rgb}{0.18,0.55,0.34}
\definecolor{SeaGreen}{rgb}{0.18,0.55,0.34}
\definecolor{SkyBlue1}{rgb}{0.53,0.81,1.00}
\definecolor{SkyBlue2}{rgb}{0.49,0.75,0.93}
\definecolor{SkyBlue3}{rgb}{0.42,0.65,0.80}
\definecolor{SkyBlue4}{rgb}{0.29,0.44,0.55}
\definecolor{SkyBlue}{rgb}{0.53,0.81,0.92}
\definecolor{SlateBlue1}{rgb}{0.51,0.44,1.00}
\definecolor{SlateBlue2}{rgb}{0.48,0.40,0.93}
\definecolor{SlateBlue3}{rgb}{0.41,0.35,0.80}
\definecolor{SlateBlue4}{rgb}{0.28,0.24,0.55}
\definecolor{SlateBlue}{rgb}{0.42,0.35,0.80}
\definecolor{SlateGray1}{rgb}{0.78,0.89,1.00}
\definecolor{SlateGray2}{rgb}{0.73,0.83,0.93}
\definecolor{SlateGray3}{rgb}{0.62,0.71,0.80}
\definecolor{SlateGray4}{rgb}{0.42,0.48,0.55}
\definecolor{SlateGray}{rgb}{0.44,0.50,0.56}
\definecolor{SlateGrey}{rgb}{0.44,0.50,0.56}
\definecolor{SpringGreen1}{rgb}{0.00,1.00,0.50}
\definecolor{SpringGreen2}{rgb}{0.00,0.93,0.46}
\definecolor{SpringGreen3}{rgb}{0.00,0.80,0.40}
\definecolor{SpringGreen4}{rgb}{0.00,0.55,0.27}
\definecolor{SpringGreen}{rgb}{0.00,1.00,0.50}
\definecolor{SteelBlue1}{rgb}{0.39,0.72,1.00}
\definecolor{SteelBlue2}{rgb}{0.36,0.67,0.93}
\definecolor{SteelBlue3}{rgb}{0.31,0.58,0.80}
\definecolor{SteelBlue4}{rgb}{0.21,0.39,0.55}
\definecolor{SteelBlue}{rgb}{0.27,0.51,0.71}
\definecolor{VioletRed1}{rgb}{1.00,0.24,0.59}
\definecolor{VioletRed2}{rgb}{0.93,0.23,0.55}
\definecolor{VioletRed3}{rgb}{0.80,0.20,0.47}
\definecolor{VioletRed4}{rgb}{0.55,0.13,0.32}
\definecolor{VioletRed}{rgb}{0.82,0.13,0.56}
\definecolor{WhiteSmoke}{rgb}{0.96,0.96,0.96}
\definecolor{YellowGreen}{rgb}{0.60,0.80,0.20}
\definecolor{aliceblue}{rgb}{0.94,0.97,1.00}
\definecolor{antiquewhite}{rgb}{0.98,0.92,0.84}
\definecolor{aquamarine1}{rgb}{0.50,1.00,0.83}
\definecolor{aquamarine2}{rgb}{0.46,0.93,0.78}
\definecolor{aquamarine3}{rgb}{0.40,0.80,0.67}
\definecolor{aquamarine4}{rgb}{0.27,0.55,0.45}
\definecolor{aquamarine}{rgb}{0.50,1.00,0.83}
\definecolor{azure1}{rgb}{0.94,1.00,1.00}
\definecolor{azure2}{rgb}{0.88,0.93,0.93}
\definecolor{azure3}{rgb}{0.76,0.80,0.80}
\definecolor{azure4}{rgb}{0.51,0.55,0.55}
\definecolor{azure}{rgb}{0.94,1.00,1.00}
\definecolor{beige}{rgb}{0.96,0.96,0.86}
\definecolor{bisque1}{rgb}{1.00,0.89,0.77}
\definecolor{bisque2}{rgb}{0.93,0.84,0.72}
\definecolor{bisque3}{rgb}{0.80,0.72,0.62}
\definecolor{bisque4}{rgb}{0.55,0.49,0.42}
\definecolor{bisque}{rgb}{1.00,0.89,0.77}
\definecolor{black}{rgb}{0.00,0.00,0.00}
\definecolor{blanchedalmond}{rgb}{1.00,0.92,0.80}
\definecolor{blue1}{rgb}{0.00,0.00,1.00}
\definecolor{blue2}{rgb}{0.00,0.00,0.93}
\definecolor{blue3}{rgb}{0.00,0.00,0.80}
\definecolor{blue4}{rgb}{0.00,0.00,0.55}
\definecolor{blueviolet}{rgb}{0.54,0.17,0.89}
\definecolor{blue}{rgb}{0.00,0.00,1.00}
\definecolor{brown1}{rgb}{1.00,0.25,0.25}
\definecolor{brown2}{rgb}{0.93,0.23,0.23}
\definecolor{brown3}{rgb}{0.80,0.20,0.20}
\definecolor{brown4}{rgb}{0.55,0.14,0.14}
\definecolor{brown}{rgb}{0.65,0.16,0.16}
\definecolor{burlywood1}{rgb}{1.00,0.83,0.61}
\definecolor{burlywood2}{rgb}{0.93,0.77,0.57}
\definecolor{burlywood3}{rgb}{0.80,0.67,0.49}
\definecolor{burlywood4}{rgb}{0.55,0.45,0.33}
\definecolor{burlywood}{rgb}{0.87,0.72,0.53}
\definecolor{cadetblue}{rgb}{0.37,0.62,0.63}
\definecolor{chartreuse1}{rgb}{0.50,1.00,0.00}
\definecolor{chartreuse2}{rgb}{0.46,0.93,0.00}
\definecolor{chartreuse3}{rgb}{0.40,0.80,0.00}
\definecolor{chartreuse4}{rgb}{0.27,0.55,0.00}
\definecolor{chartreuse}{rgb}{0.50,1.00,0.00}
\definecolor{chocolate1}{rgb}{1.00,0.50,0.14}
\definecolor{chocolate2}{rgb}{0.93,0.46,0.13}
\definecolor{chocolate3}{rgb}{0.80,0.40,0.11}
\definecolor{chocolate4}{rgb}{0.55,0.27,0.07}
\definecolor{chocolate}{rgb}{0.82,0.41,0.12}
\definecolor{coral1}{rgb}{1.00,0.45,0.34}
\definecolor{coral2}{rgb}{0.93,0.42,0.31}
\definecolor{coral3}{rgb}{0.80,0.36,0.27}
\definecolor{coral4}{rgb}{0.55,0.24,0.18}
\definecolor{coral}{rgb}{1.00,0.50,0.31}
\definecolor{cornflowerblue}{rgb}{0.39,0.58,0.93}
\definecolor{cornsilk1}{rgb}{1.00,0.97,0.86}
\definecolor{cornsilk2}{rgb}{0.93,0.91,0.80}
\definecolor{cornsilk3}{rgb}{0.80,0.78,0.69}
\definecolor{cornsilk4}{rgb}{0.55,0.53,0.47}
\definecolor{cornsilk}{rgb}{1.00,0.97,0.86}
\definecolor{cyan1}{rgb}{0.00,1.00,1.00}
\definecolor{cyan2}{rgb}{0.00,0.93,0.93}
\definecolor{cyan3}{rgb}{0.00,0.80,0.80}
\definecolor{cyan4}{rgb}{0.00,0.55,0.55}
\definecolor{cyan}{rgb}{0.00,1.00,1.00}
\definecolor{darkblue}{rgb}{0.00,0.00,0.55}
\definecolor{darkcyan}{rgb}{0.00,0.55,0.55}
\definecolor{darkgoldenrod}{rgb}{0.72,0.53,0.04}
\definecolor{darkgray}{rgb}{0.66,0.66,0.66}
\definecolor{darkgreen}{rgb}{0.00,0.39,0.00}
\definecolor{darkgrey}{rgb}{0.66,0.66,0.66}
\definecolor{darkkhaki}{rgb}{0.74,0.72,0.42}
\definecolor{darkmagenta}{rgb}{0.55,0.00,0.55}
\definecolor{darkolive}{rgb}{0.33,0.42,0.18}
\definecolor{darkorange}{rgb}{1.00,0.55,0.00}
\definecolor{darkorchid}{rgb}{0.60,0.20,0.80}
\definecolor{darkred}{rgb}{0.55,0.00,0.00}
\definecolor{darksalmon}{rgb}{0.91,0.59,0.48}
\definecolor{darksea}{rgb}{0.56,0.74,0.56}
\definecolor{darkslate}{rgb}{0.18,0.31,0.31}
\definecolor{darkslate}{rgb}{0.18,0.31,0.31}
\definecolor{darkslate}{rgb}{0.28,0.24,0.55}
\definecolor{darkturquoise}{rgb}{0.00,0.81,0.82}
\definecolor{darkviolet}{rgb}{0.58,0.00,0.83}
\definecolor{deeppink}{rgb}{1.00,0.08,0.58}
\definecolor{deepsky}{rgb}{0.00,0.75,1.00}
\definecolor{dimgray}{rgb}{0.41,0.41,0.41}
\definecolor{dimgrey}{rgb}{0.41,0.41,0.41}
\definecolor{dodgerblue}{rgb}{0.12,0.56,1.00}
\definecolor{firebrick1}{rgb}{1.00,0.19,0.19}
\definecolor{firebrick2}{rgb}{0.93,0.17,0.17}
\definecolor{firebrick3}{rgb}{0.80,0.15,0.15}
\definecolor{firebrick4}{rgb}{0.55,0.10,0.10}
\definecolor{firebrick}{rgb}{0.70,0.13,0.13}
\definecolor{floralwhite}{rgb}{1.00,0.98,0.94}
\definecolor{forestgreen}{rgb}{0.13,0.55,0.13}
\definecolor{gainsboro}{rgb}{0.86,0.86,0.86}
\definecolor{ghostwhite}{rgb}{0.97,0.97,1.00}
\definecolor{gold1}{rgb}{1.00,0.84,0.00}
\definecolor{gold2}{rgb}{0.93,0.79,0.00}
\definecolor{gold3}{rgb}{0.80,0.68,0.00}
\definecolor{gold4}{rgb}{0.55,0.46,0.00}
\definecolor{goldenrod1}{rgb}{1.00,0.76,0.15}
\definecolor{goldenrod2}{rgb}{0.93,0.71,0.13}
\definecolor{goldenrod3}{rgb}{0.80,0.61,0.11}
\definecolor{goldenrod4}{rgb}{0.55,0.41,0.08}
\definecolor{goldenrod}{rgb}{0.85,0.65,0.13}
\definecolor{gold}{rgb}{1.00,0.84,0.00}
\definecolor{gray0}{rgb}{0.00,0.00,0.00}
\definecolor{gray100}{rgb}{1.00,1.00,1.00}
\definecolor{gray10}{rgb}{0.10,0.10,0.10}
\definecolor{gray11}{rgb}{0.11,0.11,0.11}
\definecolor{gray12}{rgb}{0.12,0.12,0.12}
\definecolor{gray13}{rgb}{0.13,0.13,0.13}
\definecolor{gray14}{rgb}{0.14,0.14,0.14}
\definecolor{gray15}{rgb}{0.15,0.15,0.15}
\definecolor{gray16}{rgb}{0.16,0.16,0.16}
\definecolor{gray17}{rgb}{0.17,0.17,0.17}
\definecolor{gray18}{rgb}{0.18,0.18,0.18}
\definecolor{gray19}{rgb}{0.19,0.19,0.19}
\definecolor{gray1}{rgb}{0.01,0.01,0.01}
\definecolor{gray20}{rgb}{0.20,0.20,0.20}
\definecolor{gray21}{rgb}{0.21,0.21,0.21}
\definecolor{gray22}{rgb}{0.22,0.22,0.22}
\definecolor{gray23}{rgb}{0.23,0.23,0.23}
\definecolor{gray24}{rgb}{0.24,0.24,0.24}
\definecolor{gray25}{rgb}{0.25,0.25,0.25}
\definecolor{gray26}{rgb}{0.26,0.26,0.26}
\definecolor{gray27}{rgb}{0.27,0.27,0.27}
\definecolor{gray28}{rgb}{0.28,0.28,0.28}
\definecolor{gray29}{rgb}{0.29,0.29,0.29}
\definecolor{gray2}{rgb}{0.02,0.02,0.02}
\definecolor{gray30}{rgb}{0.30,0.30,0.30}
\definecolor{gray31}{rgb}{0.31,0.31,0.31}
\definecolor{gray32}{rgb}{0.32,0.32,0.32}
\definecolor{gray33}{rgb}{0.33,0.33,0.33}
\definecolor{gray34}{rgb}{0.34,0.34,0.34}
\definecolor{gray35}{rgb}{0.35,0.35,0.35}
\definecolor{gray36}{rgb}{0.36,0.36,0.36}
\definecolor{gray37}{rgb}{0.37,0.37,0.37}
\definecolor{gray38}{rgb}{0.38,0.38,0.38}
\definecolor{gray39}{rgb}{0.39,0.39,0.39}
\definecolor{gray3}{rgb}{0.03,0.03,0.03}
\definecolor{gray40}{rgb}{0.40,0.40,0.40}
\definecolor{gray41}{rgb}{0.41,0.41,0.41}
\definecolor{gray42}{rgb}{0.42,0.42,0.42}
\definecolor{gray43}{rgb}{0.43,0.43,0.43}
\definecolor{gray44}{rgb}{0.44,0.44,0.44}
\definecolor{gray45}{rgb}{0.45,0.45,0.45}
\definecolor{gray46}{rgb}{0.46,0.46,0.46}
\definecolor{gray47}{rgb}{0.47,0.47,0.47}
\definecolor{gray48}{rgb}{0.48,0.48,0.48}
\definecolor{gray49}{rgb}{0.49,0.49,0.49}
\definecolor{gray4}{rgb}{0.04,0.04,0.04}
\definecolor{gray50}{rgb}{0.50,0.50,0.50}
\definecolor{gray51}{rgb}{0.51,0.51,0.51}
\definecolor{gray52}{rgb}{0.52,0.52,0.52}
\definecolor{gray53}{rgb}{0.53,0.53,0.53}
\definecolor{gray54}{rgb}{0.54,0.54,0.54}
\definecolor{gray55}{rgb}{0.55,0.55,0.55}
\definecolor{gray56}{rgb}{0.56,0.56,0.56}
\definecolor{gray57}{rgb}{0.57,0.57,0.57}
\definecolor{gray58}{rgb}{0.58,0.58,0.58}
\definecolor{gray59}{rgb}{0.59,0.59,0.59}
\definecolor{gray5}{rgb}{0.05,0.05,0.05}
\definecolor{gray60}{rgb}{0.60,0.60,0.60}
\definecolor{gray61}{rgb}{0.61,0.61,0.61}
\definecolor{gray62}{rgb}{0.62,0.62,0.62}
\definecolor{gray63}{rgb}{0.63,0.63,0.63}
\definecolor{gray64}{rgb}{0.64,0.64,0.64}
\definecolor{gray65}{rgb}{0.65,0.65,0.65}
\definecolor{gray66}{rgb}{0.66,0.66,0.66}
\definecolor{gray67}{rgb}{0.67,0.67,0.67}
\definecolor{gray68}{rgb}{0.68,0.68,0.68}
\definecolor{gray69}{rgb}{0.69,0.69,0.69}
\definecolor{gray6}{rgb}{0.06,0.06,0.06}
\definecolor{gray70}{rgb}{0.70,0.70,0.70}
\definecolor{gray71}{rgb}{0.71,0.71,0.71}
\definecolor{gray72}{rgb}{0.72,0.72,0.72}
\definecolor{gray73}{rgb}{0.73,0.73,0.73}
\definecolor{gray74}{rgb}{0.74,0.74,0.74}
\definecolor{gray75}{rgb}{0.75,0.75,0.75}
\definecolor{gray76}{rgb}{0.76,0.76,0.76}
\definecolor{gray77}{rgb}{0.77,0.77,0.77}
\definecolor{gray78}{rgb}{0.78,0.78,0.78}
\definecolor{gray79}{rgb}{0.79,0.79,0.79}
\definecolor{gray7}{rgb}{0.07,0.07,0.07}
\definecolor{gray80}{rgb}{0.80,0.80,0.80}
\definecolor{gray81}{rgb}{0.81,0.81,0.81}
\definecolor{gray82}{rgb}{0.82,0.82,0.82}
\definecolor{gray83}{rgb}{0.83,0.83,0.83}
\definecolor{gray84}{rgb}{0.84,0.84,0.84}
\definecolor{gray85}{rgb}{0.85,0.85,0.85}
\definecolor{gray86}{rgb}{0.86,0.86,0.86}
\definecolor{gray87}{rgb}{0.87,0.87,0.87}
\definecolor{gray88}{rgb}{0.88,0.88,0.88}
\definecolor{gray89}{rgb}{0.89,0.89,0.89}
\definecolor{gray8}{rgb}{0.08,0.08,0.08}
\definecolor{gray90}{rgb}{0.90,0.90,0.90}
\definecolor{gray91}{rgb}{0.91,0.91,0.91}
\definecolor{gray92}{rgb}{0.92,0.92,0.92}
\definecolor{gray93}{rgb}{0.93,0.93,0.93}
\definecolor{gray94}{rgb}{0.94,0.94,0.94}
\definecolor{gray95}{rgb}{0.95,0.95,0.95}
\definecolor{gray96}{rgb}{0.96,0.96,0.96}
\definecolor{gray97}{rgb}{0.97,0.97,0.97}
\definecolor{gray98}{rgb}{0.98,0.98,0.98}
\definecolor{gray99}{rgb}{0.99,0.99,0.99}
\definecolor{gray9}{rgb}{0.09,0.09,0.09}
\definecolor{gray}{rgb}{0.75,0.75,0.75}
\definecolor{green1}{rgb}{0.00,1.00,0.00}
\definecolor{green2}{rgb}{0.00,0.93,0.00}
\definecolor{green3}{rgb}{0.00,0.80,0.00}
\definecolor{green4}{rgb}{0.00,0.55,0.00}
\definecolor{greenyellow}{rgb}{0.68,1.00,0.18}
\definecolor{green}{rgb}{0.00,1.00,0.00}
\definecolor{grey0}{rgb}{0.00,0.00,0.00}
\definecolor{grey100}{rgb}{1.00,1.00,1.00}
\definecolor{grey10}{rgb}{0.10,0.10,0.10}
\definecolor{grey11}{rgb}{0.11,0.11,0.11}
\definecolor{grey12}{rgb}{0.12,0.12,0.12}
\definecolor{grey13}{rgb}{0.13,0.13,0.13}
\definecolor{grey14}{rgb}{0.14,0.14,0.14}
\definecolor{grey15}{rgb}{0.15,0.15,0.15}
\definecolor{grey16}{rgb}{0.16,0.16,0.16}
\definecolor{grey17}{rgb}{0.17,0.17,0.17}
\definecolor{grey18}{rgb}{0.18,0.18,0.18}
\definecolor{grey19}{rgb}{0.19,0.19,0.19}
\definecolor{grey1}{rgb}{0.01,0.01,0.01}
\definecolor{grey20}{rgb}{0.20,0.20,0.20}
\definecolor{grey21}{rgb}{0.21,0.21,0.21}
\definecolor{grey22}{rgb}{0.22,0.22,0.22}
\definecolor{grey23}{rgb}{0.23,0.23,0.23}
\definecolor{grey24}{rgb}{0.24,0.24,0.24}
\definecolor{grey25}{rgb}{0.25,0.25,0.25}
\definecolor{grey26}{rgb}{0.26,0.26,0.26}
\definecolor{grey27}{rgb}{0.27,0.27,0.27}
\definecolor{grey28}{rgb}{0.28,0.28,0.28}
\definecolor{grey29}{rgb}{0.29,0.29,0.29}
\definecolor{grey2}{rgb}{0.02,0.02,0.02}
\definecolor{grey30}{rgb}{0.30,0.30,0.30}
\definecolor{grey31}{rgb}{0.31,0.31,0.31}
\definecolor{grey32}{rgb}{0.32,0.32,0.32}
\definecolor{grey33}{rgb}{0.33,0.33,0.33}
\definecolor{grey34}{rgb}{0.34,0.34,0.34}
\definecolor{grey35}{rgb}{0.35,0.35,0.35}
\definecolor{grey36}{rgb}{0.36,0.36,0.36}
\definecolor{grey37}{rgb}{0.37,0.37,0.37}
\definecolor{grey38}{rgb}{0.38,0.38,0.38}
\definecolor{grey39}{rgb}{0.39,0.39,0.39}
\definecolor{grey3}{rgb}{0.03,0.03,0.03}
\definecolor{grey40}{rgb}{0.40,0.40,0.40}
\definecolor{grey41}{rgb}{0.41,0.41,0.41}
\definecolor{grey42}{rgb}{0.42,0.42,0.42}
\definecolor{grey43}{rgb}{0.43,0.43,0.43}
\definecolor{grey44}{rgb}{0.44,0.44,0.44}
\definecolor{grey45}{rgb}{0.45,0.45,0.45}
\definecolor{grey46}{rgb}{0.46,0.46,0.46}
\definecolor{grey47}{rgb}{0.47,0.47,0.47}
\definecolor{grey48}{rgb}{0.48,0.48,0.48}
\definecolor{grey49}{rgb}{0.49,0.49,0.49}
\definecolor{grey4}{rgb}{0.04,0.04,0.04}
\definecolor{grey50}{rgb}{0.50,0.50,0.50}
\definecolor{grey51}{rgb}{0.51,0.51,0.51}
\definecolor{grey52}{rgb}{0.52,0.52,0.52}
\definecolor{grey53}{rgb}{0.53,0.53,0.53}
\definecolor{grey54}{rgb}{0.54,0.54,0.54}
\definecolor{grey55}{rgb}{0.55,0.55,0.55}
\definecolor{grey56}{rgb}{0.56,0.56,0.56}
\definecolor{grey57}{rgb}{0.57,0.57,0.57}
\definecolor{grey58}{rgb}{0.58,0.58,0.58}
\definecolor{grey59}{rgb}{0.59,0.59,0.59}
\definecolor{grey5}{rgb}{0.05,0.05,0.05}
\definecolor{grey60}{rgb}{0.60,0.60,0.60}
\definecolor{grey61}{rgb}{0.61,0.61,0.61}
\definecolor{grey62}{rgb}{0.62,0.62,0.62}
\definecolor{grey63}{rgb}{0.63,0.63,0.63}
\definecolor{grey64}{rgb}{0.64,0.64,0.64}
\definecolor{grey65}{rgb}{0.65,0.65,0.65}
\definecolor{grey66}{rgb}{0.66,0.66,0.66}
\definecolor{grey67}{rgb}{0.67,0.67,0.67}
\definecolor{grey68}{rgb}{0.68,0.68,0.68}
\definecolor{grey69}{rgb}{0.69,0.69,0.69}
\definecolor{grey6}{rgb}{0.06,0.06,0.06}
\definecolor{grey70}{rgb}{0.70,0.70,0.70}
\definecolor{grey71}{rgb}{0.71,0.71,0.71}
\definecolor{grey72}{rgb}{0.72,0.72,0.72}
\definecolor{grey73}{rgb}{0.73,0.73,0.73}
\definecolor{grey74}{rgb}{0.74,0.74,0.74}
\definecolor{grey75}{rgb}{0.75,0.75,0.75}
\definecolor{grey76}{rgb}{0.76,0.76,0.76}
\definecolor{grey77}{rgb}{0.77,0.77,0.77}
\definecolor{grey78}{rgb}{0.78,0.78,0.78}
\definecolor{grey79}{rgb}{0.79,0.79,0.79}
\definecolor{grey7}{rgb}{0.07,0.07,0.07}
\definecolor{grey80}{rgb}{0.80,0.80,0.80}
\definecolor{grey81}{rgb}{0.81,0.81,0.81}
\definecolor{grey82}{rgb}{0.82,0.82,0.82}
\definecolor{grey83}{rgb}{0.83,0.83,0.83}
\definecolor{grey84}{rgb}{0.84,0.84,0.84}
\definecolor{grey85}{rgb}{0.85,0.85,0.85}
\definecolor{grey86}{rgb}{0.86,0.86,0.86}
\definecolor{grey87}{rgb}{0.87,0.87,0.87}
\definecolor{grey88}{rgb}{0.88,0.88,0.88}
\definecolor{grey89}{rgb}{0.89,0.89,0.89}
\definecolor{grey8}{rgb}{0.08,0.08,0.08}
\definecolor{grey90}{rgb}{0.90,0.90,0.90}
\definecolor{grey91}{rgb}{0.91,0.91,0.91}
\definecolor{grey92}{rgb}{0.92,0.92,0.92}
\definecolor{grey93}{rgb}{0.93,0.93,0.93}
\definecolor{grey94}{rgb}{0.94,0.94,0.94}
\definecolor{grey95}{rgb}{0.95,0.95,0.95}
\definecolor{grey96}{rgb}{0.96,0.96,0.96}
\definecolor{grey97}{rgb}{0.97,0.97,0.97}
\definecolor{grey98}{rgb}{0.98,0.98,0.98}
\definecolor{grey99}{rgb}{0.99,0.99,0.99}
\definecolor{grey9}{rgb}{0.09,0.09,0.09}
\definecolor{grey}{rgb}{0.75,0.75,0.75}
\definecolor{honeydew1}{rgb}{0.94,1.00,0.94}
\definecolor{honeydew2}{rgb}{0.88,0.93,0.88}
\definecolor{honeydew3}{rgb}{0.76,0.80,0.76}
\definecolor{honeydew4}{rgb}{0.51,0.55,0.51}
\definecolor{honeydew}{rgb}{0.94,1.00,0.94}
\definecolor{hotpink}{rgb}{1.00,0.41,0.71}
\definecolor{indianred}{rgb}{0.80,0.36,0.36}
\definecolor{ivory1}{rgb}{1.00,1.00,0.94}
\definecolor{ivory2}{rgb}{0.93,0.93,0.88}
\definecolor{ivory3}{rgb}{0.80,0.80,0.76}
\definecolor{ivory4}{rgb}{0.55,0.55,0.51}
\definecolor{ivory}{rgb}{1.00,1.00,0.94}
\definecolor{khaki1}{rgb}{1.00,0.96,0.56}
\definecolor{khaki2}{rgb}{0.93,0.90,0.52}
\definecolor{khaki3}{rgb}{0.80,0.78,0.45}
\definecolor{khaki4}{rgb}{0.55,0.53,0.31}
\definecolor{khaki}{rgb}{0.94,0.90,0.55}
\definecolor{lavenderblush}{rgb}{1.00,0.94,0.96}
\definecolor{lavender}{rgb}{0.90,0.90,0.98}
\definecolor{lawngreen}{rgb}{0.49,0.99,0.00}
\definecolor{lemonchiffon}{rgb}{1.00,0.98,0.80}
\definecolor{lightblue}{rgb}{0.68,0.85,0.90}
\definecolor{lightcoral}{rgb}{0.94,0.50,0.50}
\definecolor{lightcyan}{rgb}{0.88,1.00,1.00}
\definecolor{lightgoldenrod}{rgb}{0.93,0.87,0.51}
\definecolor{lightgoldenrod}{rgb}{0.98,0.98,0.82}
\definecolor{lightgray}{rgb}{0.83,0.83,0.83}
\definecolor{lightgreen}{rgb}{0.56,0.93,0.56}
\definecolor{lightgrey}{rgb}{0.83,0.83,0.83}
\definecolor{lightpink}{rgb}{1.00,0.71,0.76}
\definecolor{lightsalmon}{rgb}{1.00,0.63,0.48}
\definecolor{lightsea}{rgb}{0.13,0.70,0.67}
\definecolor{lightsky}{rgb}{0.53,0.81,0.98}
\definecolor{lightslate}{rgb}{0.47,0.53,0.60}
\definecolor{lightslate}{rgb}{0.47,0.53,0.60}
\definecolor{lightslate}{rgb}{0.52,0.44,1.00}
\definecolor{lightsteel}{rgb}{0.69,0.77,0.87}
\definecolor{lightyellow}{rgb}{1.00,1.00,0.88}
\definecolor{limegreen}{rgb}{0.20,0.80,0.20}
\definecolor{linen}{rgb}{0.98,0.94,0.90}
\definecolor{magenta1}{rgb}{1.00,0.00,1.00}
\definecolor{magenta2}{rgb}{0.93,0.00,0.93}
\definecolor{magenta3}{rgb}{0.80,0.00,0.80}
\definecolor{magenta4}{rgb}{0.55,0.00,0.55}
\definecolor{magenta}{rgb}{1.00,0.00,1.00}
\definecolor{maroon1}{rgb}{1.00,0.20,0.70}
\definecolor{maroon2}{rgb}{0.93,0.19,0.65}
\definecolor{maroon3}{rgb}{0.80,0.16,0.56}
\definecolor{maroon4}{rgb}{0.55,0.11,0.38}
\definecolor{maroon}{rgb}{0.69,0.19,0.38}
\definecolor{mediumaquamarine}{rgb}{0.40,0.80,0.67}
\definecolor{mediumblue}{rgb}{0.00,0.00,0.80}
\definecolor{mediumorchid}{rgb}{0.73,0.33,0.83}
\definecolor{mediumpurple}{rgb}{0.58,0.44,0.86}
\definecolor{mediumsea}{rgb}{0.24,0.70,0.44}
\definecolor{mediumslate}{rgb}{0.48,0.41,0.93}
\definecolor{mediumspring}{rgb}{0.00,0.98,0.60}
\definecolor{mediumturquoise}{rgb}{0.28,0.82,0.80}
\definecolor{mediumviolet}{rgb}{0.78,0.08,0.52}
\definecolor{midnightblue}{rgb}{0.10,0.10,0.44}
\definecolor{mintcream}{rgb}{0.96,1.00,0.98}
\definecolor{mistyrose}{rgb}{1.00,0.89,0.88}
\definecolor{moccasin}{rgb}{1.00,0.89,0.71}
\definecolor{navajowhite}{rgb}{1.00,0.87,0.68}
\definecolor{navyblue}{rgb}{0.00,0.00,0.50}
\definecolor{navy}{rgb}{0.00,0.00,0.50}
\definecolor{oldlace}{rgb}{0.99,0.96,0.90}
\definecolor{olivedrab}{rgb}{0.42,0.56,0.14}
\definecolor{orange1}{rgb}{1.00,0.65,0.00}
\definecolor{orange2}{rgb}{0.93,0.60,0.00}
\definecolor{orange3}{rgb}{0.80,0.52,0.00}
\definecolor{orange4}{rgb}{0.55,0.35,0.00}
\definecolor{orangered}{rgb}{1.00,0.27,0.00}
\definecolor{orange}{rgb}{1.00,0.65,0.00}
\definecolor{orchid1}{rgb}{1.00,0.51,0.98}
\definecolor{orchid2}{rgb}{0.93,0.48,0.91}
\definecolor{orchid3}{rgb}{0.80,0.41,0.79}
\definecolor{orchid4}{rgb}{0.55,0.28,0.54}
\definecolor{orchid}{rgb}{0.85,0.44,0.84}
\definecolor{palegoldenrod}{rgb}{0.93,0.91,0.67}
\definecolor{palegreen}{rgb}{0.60,0.98,0.60}
\definecolor{paleturquoise}{rgb}{0.69,0.93,0.93}
\definecolor{paleviolet}{rgb}{0.86,0.44,0.58}
\definecolor{papayawhip}{rgb}{1.00,0.94,0.84}
\definecolor{peachpuff}{rgb}{1.00,0.85,0.73}
\definecolor{peru}{rgb}{0.80,0.52,0.25}
\definecolor{pink1}{rgb}{1.00,0.71,0.77}
\definecolor{pink2}{rgb}{0.93,0.66,0.72}
\definecolor{pink3}{rgb}{0.80,0.57,0.62}
\definecolor{pink4}{rgb}{0.55,0.39,0.42}
\definecolor{pink}{rgb}{1.00,0.75,0.80}
\definecolor{plum1}{rgb}{1.00,0.73,1.00}
\definecolor{plum2}{rgb}{0.93,0.68,0.93}
\definecolor{plum3}{rgb}{0.80,0.59,0.80}
\definecolor{plum4}{rgb}{0.55,0.40,0.55}
\definecolor{plum}{rgb}{0.87,0.63,0.87}
\definecolor{powderblue}{rgb}{0.69,0.88,0.90}
\definecolor{purple1}{rgb}{0.61,0.19,1.00}
\definecolor{purple2}{rgb}{0.57,0.17,0.93}
\definecolor{purple3}{rgb}{0.49,0.15,0.80}
\definecolor{purple4}{rgb}{0.33,0.10,0.55}
\definecolor{purple}{rgb}{0.63,0.13,0.94}
\definecolor{red1}{rgb}{1.00,0.00,0.00}
\definecolor{red2}{rgb}{0.93,0.00,0.00}
\definecolor{red3}{rgb}{0.80,0.00,0.00}
\definecolor{red4}{rgb}{0.55,0.00,0.00}
\definecolor{red}{rgb}{1.00,0.00,0.00}
\definecolor{rosybrown}{rgb}{0.74,0.56,0.56}
\definecolor{royalblue}{rgb}{0.25,0.41,0.88}
\definecolor{saddlebrown}{rgb}{0.55,0.27,0.07}
\definecolor{salmon1}{rgb}{1.00,0.55,0.41}
\definecolor{salmon2}{rgb}{0.93,0.51,0.38}
\definecolor{salmon3}{rgb}{0.80,0.44,0.33}
\definecolor{salmon4}{rgb}{0.55,0.30,0.22}
\definecolor{salmon}{rgb}{0.98,0.50,0.45}
\definecolor{sandybrown}{rgb}{0.96,0.64,0.38}
\definecolor{seagreen}{rgb}{0.18,0.55,0.34}
\definecolor{seashell1}{rgb}{1.00,0.96,0.93}
\definecolor{seashell2}{rgb}{0.93,0.90,0.87}
\definecolor{seashell3}{rgb}{0.80,0.77,0.75}
\definecolor{seashell4}{rgb}{0.55,0.53,0.51}
\definecolor{seashell}{rgb}{1.00,0.96,0.93}
\definecolor{sienna1}{rgb}{1.00,0.51,0.28}
\definecolor{sienna2}{rgb}{0.93,0.47,0.26}
\definecolor{sienna3}{rgb}{0.80,0.41,0.22}
\definecolor{sienna4}{rgb}{0.55,0.28,0.15}
\definecolor{sienna}{rgb}{0.63,0.32,0.18}
\definecolor{skyblue}{rgb}{0.53,0.81,0.92}
\definecolor{slateblue}{rgb}{0.42,0.35,0.80}
\definecolor{slategray}{rgb}{0.44,0.50,0.56}
\definecolor{slategrey}{rgb}{0.44,0.50,0.56}
\definecolor{snow1}{rgb}{1.00,0.98,0.98}
\definecolor{snow2}{rgb}{0.93,0.91,0.91}
\definecolor{snow3}{rgb}{0.80,0.79,0.79}
\definecolor{snow4}{rgb}{0.55,0.54,0.54}
\definecolor{snow}{rgb}{1.00,0.98,0.98}
\definecolor{springgreen}{rgb}{0.00,1.00,0.50}
\definecolor{steelblue}{rgb}{0.27,0.51,0.71}
\definecolor{tan1}{rgb}{1.00,0.65,0.31}
\definecolor{tan2}{rgb}{0.93,0.60,0.29}
\definecolor{tan3}{rgb}{0.80,0.52,0.25}
\definecolor{tan4}{rgb}{0.55,0.35,0.17}
\definecolor{tan}{rgb}{0.82,0.71,0.55}
\definecolor{thistle1}{rgb}{1.00,0.88,1.00}
\definecolor{thistle2}{rgb}{0.93,0.82,0.93}
\definecolor{thistle3}{rgb}{0.80,0.71,0.80}
\definecolor{thistle4}{rgb}{0.55,0.48,0.55}
\definecolor{thistle}{rgb}{0.85,0.75,0.85}
\definecolor{tomato1}{rgb}{1.00,0.39,0.28}
\definecolor{tomato2}{rgb}{0.93,0.36,0.26}
\definecolor{tomato3}{rgb}{0.80,0.31,0.22}
\definecolor{tomato4}{rgb}{0.55,0.21,0.15}
\definecolor{tomato}{rgb}{1.00,0.39,0.28}
\definecolor{turquoise1}{rgb}{0.00,0.96,1.00}
\definecolor{turquoise2}{rgb}{0.00,0.90,0.93}
\definecolor{turquoise3}{rgb}{0.00,0.77,0.80}
\definecolor{turquoise4}{rgb}{0.00,0.53,0.55}
\definecolor{turquoise}{rgb}{0.25,0.88,0.82}
\definecolor{violetred}{rgb}{0.82,0.13,0.56}
\definecolor{violet}{rgb}{0.93,0.51,0.93}
\definecolor{wheat1}{rgb}{1.00,0.91,0.73}
\definecolor{wheat2}{rgb}{0.93,0.85,0.68}
\definecolor{wheat3}{rgb}{0.80,0.73,0.59}
\definecolor{wheat4}{rgb}{0.55,0.49,0.40}
\definecolor{wheat}{rgb}{0.96,0.87,0.70}
\definecolor{whitesmoke}{rgb}{0.96,0.96,0.96}
\definecolor{white}{rgb}{1.00,1.00,1.00}
\definecolor{yellow1}{rgb}{1.00,1.00,0.00}
\definecolor{yellow2}{rgb}{0.93,0.93,0.00}
\definecolor{yellow3}{rgb}{0.80,0.80,0.00}
\definecolor{yellow4}{rgb}{0.55,0.55,0.00}
\definecolor{yellowgreen}{rgb}{0.60,0.80,0.20}
\definecolor{yellow}{rgb}{1.00,1.00,0.00}

\newcommand{\kms}{km~s$^{-1}$}
\newcommand{\spose}[1]{\hbox to 0pt{#1\hss}}
\newcommand{\simlt}{\mathrel{\spose{\lower 3pt\hbox{$\mathchar"218$}}
     \raise 1.5pt\hbox{$\mathchar"13C$}}}
\newcommand{\simgt}{\mathrel{\spose{\lower 3pt\hbox{$\mathchar"218$}}
     \raise 1.5pt\hbox{$\mathchar"13E$}}}
\newcommand{\ion}[2]{\ifmmode 
 \mbox{\rm #1\,{\small\uppercase\expandafter{\romannumeral #2}}}
 \else\mbox{#1\,{\small\uppercase\expandafter{\romannumeral #2}}}\fi}

\begin{document}

\title[warm neutral gas towards Q$2206-199$]{The kinetic temperature in a damped Lyman-alpha absorption system in Q2206{\,}--{\,}199 - an example of the warm neutral medium}

\author[R. F. Carswell et al.]
{R. F. Carswell $^{1}$\thanks{E-mail:rfc@ast.cam.ac.uk}, G. D. Becker$^2$, R. A. Jorgenson$^{1,3}$, M. T. Murphy$^{4}$\vspace*{2.0mm} \\
{\LARGE\phantom{mm}\rm \& A. M. Wolfe$^{5}$}\\
$^{1}$Institute of Astronomy, University of Cambridge, Madingley Road, Cambridge CB3 0HA, UK\\
$^{2}$Kavli Institute for Cosmology \& Institute of Astronomy, University of Cambridge, Madingley Road, Cambridge CB3 0HA, UK\\
$^{3}$Institute for Astronomy, University of Hawaii, 2680 Woodlawn Drive, Honolulu, HI 96822, USA\\
$^{4}$Centre for Astrophysics and Supercomputing, Swinburne University of Technology, Melbourne, Victoria 3122, Australia\\
$^{5}$Center for Astrophysics \& Space Sciences, University of California, San Diego, 9500 Gilman Drive, La Jolla, CA 92093-0424, USA\\
}


\date{Accepted \phantom{mmmmmmm}. Received \phantom{mmmmmmm}; in original form \phantom{mmmmmmm}}

\pagerange{\pageref{firstpage}--\pageref{lastpage}} \pubyear{2011}

\maketitle

\label{firstpage}

\begin{abstract}
By comparing the widths of absorption lines from \ion{O}{1}, 
\ion{Si}{2} and \ion{Fe}{2} in the redshift $z_{\rm abs}=2.076$ 
single-component damped Ly$\alpha$ system in the spectrum of Q$2206-199$ 
we establish that these absorption lines arise in Warm Neutral Medium gas 
at $\sim 12000 \pm 3000$\,K. This is consistent with thermal equilibrium model 
estimates of $\sim 8000$\,K for the Warm Neutral Medium in galaxies, but not 
with the presence of a significant cold component. It is also consistent with, but 
not required by, the absence of \ion{C}{2} fine structure absorption in this system.
Some possible implications concerning abundance estimates  in narrow-line 
WNM absorbers are discussed.
\end{abstract}


\section{Introduction}

{\color{black} The complex structure of the interstellar medium in the Galaxy may be considered as a
mixture of four components: the neutral atomic hydrogen is found to be generally in  cold neutral medium (CNM, $\simlt 500$\,K) and warm neutral medium 
(WNM, $\sim 5000$ - $10000$\,K) gas \citep{Wol03}, and ionized material may be warm (WIM, $\sim 10^4$\,K) or
hot (HIM, $T\sim 10^6$\,K) \citep{McK77}. The WNM and CNM are expected to be in approximate pressure balance (e.g. Wolfire et al., 1995); \ion{H}{1} at temperatures between the two is unstable
and will rapidly evolve towards one of the two stable states. The CNM and WNM are usually studied
using \ion{H}{1} 21cm absorption against background radio sources, and show that a significant fraction of the warm material may be in the unstable regime (see the discussion by Wolfire et al., 1995, also Begum et al., 2010).}

Temperature estimates for Damped Lyman-alpha absorption systems (DLAs) reveal 
both CNM and WNM gas. The presence of \ion{C}{2} fine structure 
absorption is associated with the CNM \citep{Wol03,Wol03b}, and \citet{How05}
use \ion{C}{2} and \ion{Si}{2} fine structure lines to show that a system at 
redshift $z=4.2$
contains CNM material. \citet{Leh08} use an upper limit to the 
\ion{C}{2}$^*$ column density
in the $z=2.377$ DLA to show that the neutral gas in this system must be warm.
Modelling of a few weak \ion{Mg}{2} absorbers also suggests temperatures 
of $\sim 10000$ K \citep{Jon10}. From 21cm spin
temperature studies  \citet{Kan09} suggest that a significant fraction of 
the \ion{H}{1} at redshifts $z>1.6$ is warm. However, this conclusion is 
controversial as it requires
assumptions about the temperature and density distribution
of the gas across the ~100 pc dimensions of the background
radio sources that have not been demonstrated \citep{Wol11}.

The few attempts to estimate temperatures from the line widths in DLAs have 
relied on the identification
of narrow absorption components, normally from \ion{C}{1}, which show up 
clearly as distinct sharp features 
\citep{Jor09, Tum10, Car11}. The corresponding temperatures of a few hundred
degrees Kelvin or less indicate that the absorbing gas is in a cold neutral 
medium (CNM). 
However, attempts to determine temperatures by using the widths of singly ionized 
heavy element lines have not been very successful, either in the 
Galaxy \citep{Neh08} or at higher redshifts \citep{Car11} because 
of uncertainties introduced by the blending of many velocity components. One exception
is for the interstellar medium near the Sun, where \citet{Red04} find temperatures averaging
6800K with a dispersion of $\sim 1500$ K. 

One way of avoiding velocity component confusion is to select systems
for which the velocity structure in the singly ionized species is simple. 
These are rare, and may not be typical, but it is of interest to see if the inferred 
temperature in such cases is $\sim 8000$\,K, as would be expected for the WNM \citep{Wol95},
or if a temperature of $\simlt 500$\,K typical of the CNM is more appropriate.
The other requirements are that at least two ion species with different
atomic masses and transitions with a range of oscillator strengths be measured,
and that at least some of these fall on different parts of the curve of growth
(normally linear and logarithmic) so that the Doppler parameters for the ions 
may be estimated individually \citep{Jor09, Car11}. 

{\color{black} There are two studies where the inferred line widths have yielded 
temperature estimates in QSO absorbers dominated by single-component systems
with strong \ion{H}{1}, both as part of D/H
abundance studies. \citet{OMe01} use Keck HIRES spectra to derive a temperature 
$T=1.15\pm 0.02\times 10^4$\,K, and a bulk 
motion component $b_{\rm turb}=2.56\pm 0.12$\,{\kms} by comparing the Doppler 
parameters for the \ion{H}{1}, \ion{N}{1} and \ion{O}{1} lines in a system with 
\ion{H}{1} column density $\log N({\rm\ion{H}{1}})=19.42$ at redshift $z=2.536$ 
towards the QSO HS$0105+1619$. The other is a DLA at redshift $z=2.076$ in the 
spectrum of Q$2206-199$ which was studied by \citet{Pro97} using Keck HIRES data. 
\citet{Pet01} used a combination 
of HST STIS and VLT UVES data to derive a similar, if less certain, temperature by 
comparing the Doppler parameters for \ion{H}{1} and singly ionized heavy element lines.

We revisit the second case here, since the temperature is based largely on low S/N
STIS spectra for higher order Lyman lines. With the presence of several transitions of 
a range of ions of different masses, \ion{O}{1}, \ion{Si}{2} and \ion{Fe}{2} there is the prospect of 
temperature estimation from the heavy element lines alone. In the next section we describe 
the spectroscopic data and the techniques used to establish a temperature 
$T\sim 12000\pm 3000$\,K. Section \ref{ss:systematics} describes a series of tests
undertaken to ensure that the result is reliable, and in particular that the error range is realistic
and that the absorbing medium must be warm, and some further temperature indicators are
considered in Section \ref{ss:STIScomp}. Section \ref{sec:tempimp} examines some implications
of the temperature estimate, particularly for relative element abundance estimates in cases
where the absorption lines are intrinsically narrow, and the main conclusions are summarized
in Section \ref{sec:concl}.} 

\section{The redshift 2.076 absorption system in Q2206\,{--}\,199}
\label{sec:2206}

\subsection{The data}
\label{ss:data}

{\color{black}The absorption system at redshift $z=2.076$ in Q$2206-199$ was found to have low heavy element abundances, with  ${\rm [Si/H]} \simlt -2.4$, by \citet{Rau90} using 30\,{\kms} resolution spectra.  Using higher resolution Keck HIRES spectra,  \citet{Pro97}}  noted
the very low \ion{Fe}{2} and \ion{Al}{2} abundances in a system at redshift $z=2.076214$,
and that the system appeared to have only a single velocity component. 
\citet{Pro01} re-examined this system and further determined the abundance for \ion{Si}{2},
further confirming it as a low abundance system with ${\rm [Fe/H]}=-2.6$, and ${\rm [Si/H]}=-2.3$. 

\begin{figure*}
\includegraphics[width=120mm,angle=0]{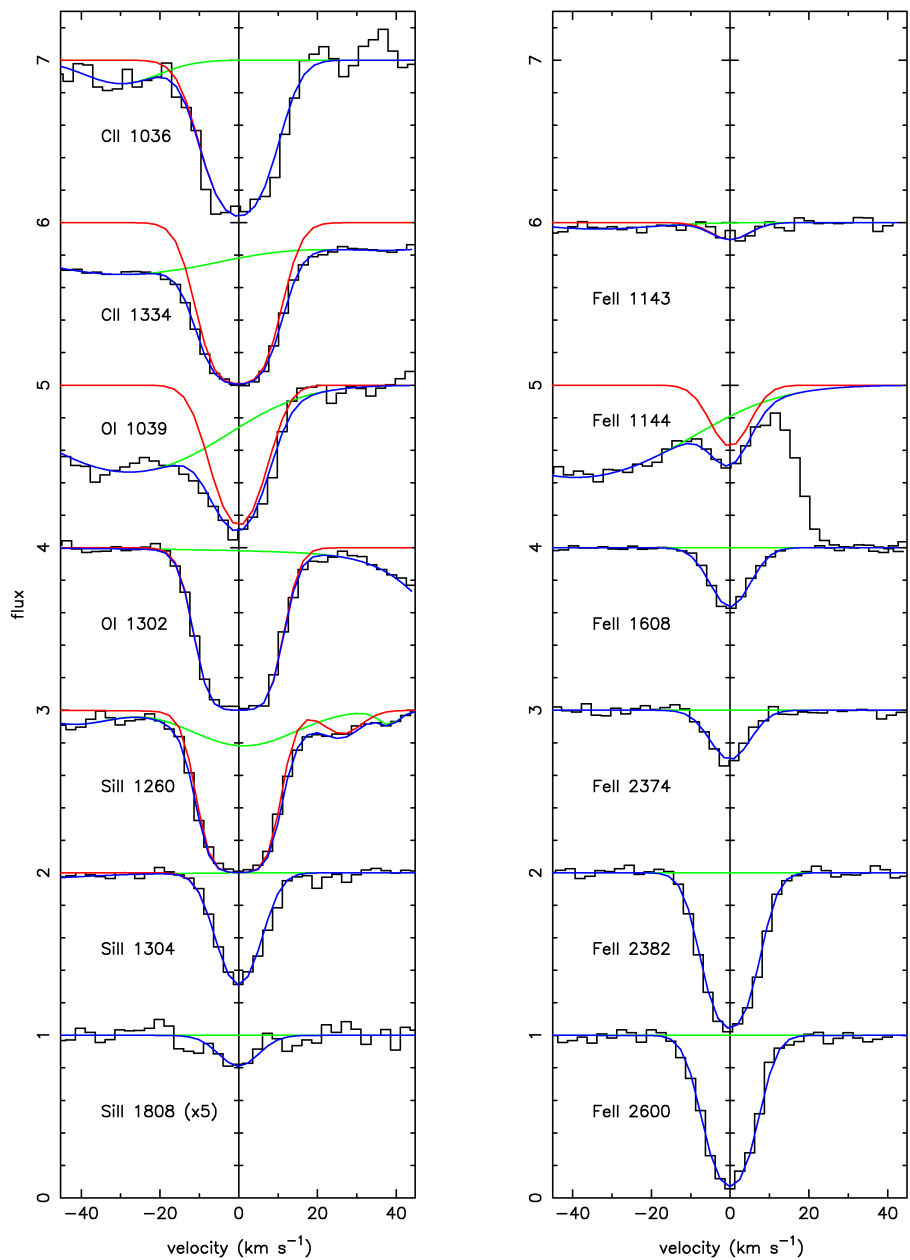}
\caption{The Voigt profile fits and data for the \ion{C}{2}, \ion{O}{1}, \ion{Si}{2} \& 
\ion{Fe}{2} lines used {\color{black}on a velocity scale relative to redshift $z=2.0762286$}. 
The data are shown as black histograms and the overall
fits in blue. Where they differ from the overall fitted profiles, the fits to the 
transitions of interest here are indicated are shown in red, and any interlopers
shown in green. All but \ion{Si}{2} 1808 are shown against a unit continuum, 
and the continua for different lines have been separated by an integer offset. 
The \ion{Si}{2} 1808 line is weak, and has been scaled up by a 
factor of five so it can be seen more clearly. The dip (in red) at $+26$\,{\kms} relative 
to the \ion{Si}{2} 1260 line is \ion{Fe}{2} 1260 at the same redshift.}
\label{fig:linprof}
\end{figure*}


The spectra used here were obtained using UVES on the ESO VLT, for programmes
65.O-0158(A) \citep{Pet02}, 072.A-0346(A) \citep{Pet06} and 074.A-0201(A). {\color{black} For each exposure ESO
provides a seeing estimate from their differential image motion monitor.} The standard 
setting DIC2 
437 (3730 - 4990\,{\AA}) \& 860 (6650 - 10600\,{\AA}) was used for two exposures 
for a total of 7800s in 0.5 arcsec seeing as part of programme 65.O-0158(A). 
The other settings used in that programme were DIC1 390 (3260 - 4450\,{\AA}) \& 
564 (4580 - 6680\,{\AA}), for 10800s 
with 0.5 arcsec seeing and 6300s with slit-limited (1 arcsec) resolution.
For programme 072.A-0346(A) the seeing approximately matched the slit width 
of 1 arcsec, and the exposures totalled 9000s for DIC1 346 
(3030 - 3880\,{\AA})  \& 580 (4760 - 6840\,{\AA}) settings. Programme 074.A-0201(A)
used a 0.9 arcsec slit in seeing ranging from 0.8 to 1.1 arcsec with total 
exposure times of 19603s in settings DIC2 455 (3930 - 5160\,{\AA}) \& 
850 (6580 - 10420\,{\AA}). The data were
reduced using the UVES pipeline and the UVES\_popler 
package\footnote{http://astronomy.swin.edu.au/$\sim$mmurphy/UVES\_popler.html}. 
The  signal-to-noise ratio (S/N) in the coadded spectrum is 60 per 2.5 
{\kms} pixel at 5000\,{\AA}. 

\begin{table*}
\centering
 \begin{minipage}{140mm}
  \caption{{\color{black}Q$2206-199$ $z=2.076$ system profile fit parameters.}\label{tab:sife}}
  \begin{tabular}{@{}lcccccccccc@{}}
  \hline
Ion&  $z$ & $\pm$& $b$ & $\pm$ & $\log(N)$&$\pm$&$b_{\rm turb}$&$\pm$&$T$\,(K)&$\pm$\\ \hline
C II&2.0762286&0.0000005&7.66&0.26&14.24&0.03\\
O I &   &    &6.43&(0.16)&15.08&0.05&5.3&0.2&$1.22\times 10^4$&$3.2\times 10^3$\\
SiII&&&5.99&(0.14)&13.69&0.01\\ 
FeII&&&5.68&(0.11)&13.35&0.01\\
\\ 
??&2.1898787&0.0000107& 2.23& 2.28&11.67&0.14\\ 
H I&1.6221834&0.0000260&12.22& 4.92&12.43&0.15\\ 
H I&1.6290246&0.0000890&26.68&12.25&12.69&0.22\\ 
H I&1.6295216&0.0000189&28.39& 4.02&13.46&0.07\\ 
H I&1.8925871&0.0000713&18.38&15.79&12.00&0.55\\ 
H I&1.8968634&0.0000063&33.22& 3.24&13.57&0.07\\ 
H I&2.1890279&0.0000208&12.42& 3.05&12.18&0.09\\ 
H I&2.1894871&0.0000125&17.10& 3.12&12.76&0.10\\ 
H I&2.2958922&0.0000256&10.88& 3.14&12.85&0.20\\ 
H I&2.2959933&0.0000046&24.56& 0.55&13.85&0.02\\ 
H I&2.2968227&0.0000682&98.83&10.40&13.46&0.06\\ 
H I&2.2977042&0.0000065&20.64& 0.53&13.95&0.02\\ 
H I&2.2980854&0.0000359&18.28& 2.54&12.97&0.11\\ 
H I&2.2998674&0.0000571&35.49&11.59&12.43&0.21\\ 
H I&2.3766727&0.0000325&39.47& 4.17&13.30&0.05\\ 
H I&2.3774852&0.0000461&31.52& 8.53&12.85&0.13\\ 
 \hline
\end{tabular}

{Column densities $N$ have units cm$^{-2}$, and Doppler parameters $b$ are in {\kms}. \par
`??' denotes an unidentified line, taken to have the same rest wavelength and oscillator strength as Ly$\alpha$.}
\end{minipage}
\end{table*}

The variety of slit widths and wavelength settings is likely to result in 
the spectral resolution being a function of wavelength. {\color{black} For individual exposures
appropriate width Gaussians provide an adequate approximation to the instrument profile.
For the slit-limited cases, which dominate the final sum here, we verified this assumption by establishing that single Gaussians 
gave good fits to night sky emission lines in the sky spectrum described by \citet{Han03}.
For the seeing-limited observations there may be extended wings beyond a core which is 
reasonably close to a Gaussian (see e.g. Diego 1985, King 1971).} 
We have constructed
an instrument profile as a sum of Gaussians with a common centre and with
widths determined using the resolution - slit width
product $R\times s=41400$ ($s$ in arcsec) for the blue arm and 
$R\times s=38700$
for the red arm of the spectrograph, {\color{black} weighting each by the relative efficiency (see the ESO UVES instrument characteristics webpage\footnote{http://www.eso.org/sci/facilities/paranal/instruments/uves/inst/} and the response curves linked to from there)}. 
For the 0.5 arcsec seeing-limited exposures we took $s$ to be the seeing 
estimate. For the other exposures the resolution was close to being 
slit-limited, so $s$ was set to the appropriate slit width.  {\color{black} While for the seeing limited exposures the adopted profile will not give the full description, any flux outside the core will
be dominated by the slit-limited observations so should not be important. In any case, as we
show below, the results here are insensitive to the line-spread-function adopted so a
precise one is not needed.}

{\color{black} The error estimates provided by the data extraction package were found to be 
close to those inferred by examining the root-mean-square (RMS) deviations in continuum
regions. However, in lines which have zero residual intensity the RMS measures indicated that 
the error estimates were too low by a factor of 2. We verified that this is generally true for UVES
spectra extracted in this way by examining data for other objects. To correct for the error underestimate when the data is near zero we multiplied the error term by 
$\left(1.0+15.0\times  {\rm min}\left(1.0,{\rm max}\left(0.0,(1.0-d_i/c_i)\right)^4\right)\right)^{1\over 4}$, 
where $d_i$, $c_i$ are the data and continuum values in the $i$th data pixel. While it is not clear what the true correction should be, this function was chosen because it changes the error little when the data is not close to zero and doubles it near zero intensity. Provided that the correction
is sharply peaked near the data zero, the precise form of the correction makes little difference to the results here.
}

\subsection{Temperature estimation}
\label{ss:temp}

The full optical wavelength coverage of these spectra allows us to
measure several \ion{Fe}{2} and \ion{Si}{2} lines with a 
range of oscillator strengths $f$ and wavelengths $\lambda$, 
and \ion{O}{1} 1039 and 1302 with $f\lambda$ values differing by 0.8 dex. So,
at least in principle, we may estimate the Doppler parameters for these 
species independently. Also, with a range of $f\lambda$ values the profile 
fits depend most strongly on the curve of growth, and not on an estimate of 
the instrument profile \citep{Car11}.  We assume that all ions arise predominantly in the same
region where a single temperature is appropriate. Since they arise in a DLA these
assumptions should be close to reality. Within the region we assume that 
the Doppler parameter for a given ion is $b=\sqrt{b_{\rm turb}^2 +
2kT/m}$, where $b_{\rm turb}$ is the turbulent (bulk) component, $k$
Boltzmann's constant, $m$ the ion mass and $T$ the temperature. We used
the VPFIT program\footnote{http://www.ast.cam.ac.uk/$\sim$rfc/vpfit.html} 
with oscillator strengths from the compilation by \citet{Mor03}. For the Doppler parameters
$b_{\rm turb}$ and $T$ were used as the independent variables for
at least two ions of different mass at the same redshift
simultaneously, with the constraints that both variables
are non-negative. The error estimates from the program then apply to
$b_{\rm turb}$ and $T$, not to the Doppler parameters for the individual ions.
The temperature component of the Doppler width varies as $1/\sqrt{m}$, so even
for \ion{O}{1} and \ion{Fe}{2} the thermal widths differ by less than a factor of two
and for other available ions the ratio is even smaller.
If there is a significant turbulent component then the differences in the
line widths could be quite small. Under these circumstances, even quite well-determined
Doppler parameters for individual ions can yield large error estimates for 
both $b_{\rm turb}$ and $T$. The Doppler parameter error estimates shown in parentheses in 
Table \ref{tab:sife} are given as a guide only, and were obtained by assuming that 
the $b$-value for each ion was determined independently of the others. 

The redshifts of \ion{C}{2} 1036 and 1334 were tied to be the same as for the other ions
to provide a (small) increase in the precision of the $z$-estimate.
However, since the \ion{C}{2}  $f\lambda$ values differ by only 0.14 dex they were not used for constraining the temperature.
The \ion{N}{1} lines at 1200\,{\AA} are too weak to be useful. 

In the $z=2.076$ absorption system \ion{Si}{2} 1260, 1304 and 1808 are 
reasonably clear of contaminating absorption lines from systems at other redshifts, but \ion{Si}{2} 1526, which is clear of the Ly$\alpha$ forest, is unusable
because it is badly blended with a strong complex \ion{Fe}{2} 1608 at redshift $z=1.9205$. Allowance has to be made for
blended Ly$\alpha$ forest lines for all usable \ion{Si}{2} lines except 1808. This is straightforward generally, since the Ly$\alpha$ lines are significantly 
broader than the heavy element ones. Some of the longer wavelength 
\ion{Fe}{2} lines are
affected by atmospheric absorption and emission, but 
\ion{Fe}{2} 1608, 2374, 2382 and 2600 are 
clear of serious contamination. At shorter wavelengths \ion{Fe}{2} 1260 
is in the wing of \ion{Si}{2} 1260, and \ion{Fe}{2} 1143 and 1144 provide 
useful constraints. For \ion{O}{1} 1302 and 1039 are also useful, though 
the 1039 line is blended with a broad Ly$\alpha$.
The results of fitting these with VPFIT are given in Table \ref{tab:sife}, along with the blends 
fitted at the same time. The Doppler
parameters for \ion{O}{1}, \ion{Si}{2} and \ion{Fe}{2} were determined using 
common values of $b_{\rm turb}$ and $T$, which were treated as independent variables. 
The error estimates for the 
Doppler parameters given in brackets are derived assuming that
they are independent of those for the other ions.  The line designated by `??', which is
assigned the same wavelength and oscillator strength as Ly$\alpha$, is
a marginal narrow component which is probably an unidentified heavy element.

{\color{black} There are two data uncertainties which should be allowed for.  The continuum level cannot be determined precisely, especially in the Ly$\alpha$ forest, and inaccurate continuum levels can cause inconsistencies when several lines of the same ion are fitted simultaneously. The other is the true zero level for the data, which can affect strong lines particularly, can differ from the data extraction zero by 1 or 2\%. The VPFIT package allows both of these to be used as
further free parameters in the overall fit, and we have done so where necessary. We do not
give their values in Table \ref{tab:sife} since they depend on how well the continuum and zero 
levels have been estimated originally.}

From Table \ref{tab:sife} we see that the temperature estimate for the system
is $T={\color{black}1.22} \pm 0.32\times 10^4$\,K, so consistent with it being WNM but not CNM. 
The turbulent component of 5.3 {\kms} is 
larger than the thermal broadening for any of the ions at the inferred temperature.
For \ion{O}{1} the thermal width is $b={\color{black} 3.5}$ {\kms}, for \ion{Si}{2} {\color{black}2.7} {\kms} and for 
\ion{Fe}{2} {\color{black}1.9} {\kms}. However, these thermal widths are enough to give measurable 
differences in the Doppler parameters for the ions involved.

\begin{table}
  \caption{Other $z=2.076$ system ions \label{tab:CAl}}
  \raggedright
  \leavevmode
  \begin{tabular}{@{}lccrcrc@{}}
  \hline
Ion&  $z$ & $\pm$& $b$\phantom{m} & $\pm$ & $\log(N)$&$\pm$\\ \hline
 H I   & 2.0762286&(0.0000005)&   15.19  & -  & 20.44&0.05\\
 C II  & 2.0762286&                      &    6.75   &     & 14.35&0.02\\
 C II$^*$& 2.0762286&                &    6.75  &     & $<12.75$ & ($2\sigma$)\\
 AlII  & 2.0762286&                       &    6.01   &     & 12.17&0.01\\
 AlIII & 2.0762286&                       &    6.01   &     & 11.46&0.06\\  
 \\
 C IV & 2.0753686&   0.0000121&     7.54&      1.88&  12.09&  0.08\\
 C IV & 2.0761144&   0.0000058&     5.11&      1.49&  12.67&  0.20\\
 C IV & 2.0762127&   0.0000286&   17.32&      2.09&  13.57&  0.04\\
 C IV & 2.0764230&   0.0000136&     9.57&      1.87&  13.00&  0.22\\
 C IV & 2.0766579&   0.0000409&   12.42&      5.11&  12.24&  0.17\\
 SiIV & 2.0761218&   0.0000020&     7.75&      0.33&  12.64&    0.02\\ 
 SiIV & 2.0764017&   0.0000065&    12.91&      1.05&  12.55&    0.03\\
\hline
\end{tabular}
Note: \ion{H}{1}, \ion{C}{2}, \ion{Al}{2} and \ion{Al}{3} redshifts and $b$-values 
are obtained from the parameters given in Table \ref{tab:sife}. 
\end{table}

The \ion{C}{2} column density given in Table \ref{tab:sife} is the best fit obtained by leaving its 
Doppler $b$ to be determined independently of the other ions. If we impose the 
turbulent component and temperature derived above, then for \ion{C}{2} {\color{black}$b=6.75$\,{\kms} 
and $\log N=14.38\pm 0.03$ }(see Table \ref{tab:CAl}). 

{\color{black}For the fitted profiles given in Table \ref{tab:sife} the reduced $\chi^2=1.103$ ($\chi^2=893.39$ for 810 degrees of freedom). The formal probability that this fit describes the data
is 0.022. One might improve the fit by adding more blended lines, but 
the the reduction in $\chi^2$ is to some extent offset by the loss in the number of degrees 
of freedom - see \citet{Rau92} for a discussion on this point. We have not quite reached that 
stage: the temperature estimate does
depend on having identified all the narrow ($< 10$\,{\kms}) blends with key lines. For example,
adding an extra narrow component to \ion{Si}{2} 1304 gives a system 
temperature of 11800\,K. However,
the \ion{Si}{2} 1260 profile suggests that this narrow component is not \ion{Si}{2}, and 
we have not found
any other lines in a system which would plausibly identify such a component.}

For this system the \ion{H}{1} column density is $\log N({\rm\ion{H}{1}})=20.43\pm 0.06$
\citep{Pet94,Pro01}. \citet{Pet94} show a fit to the Ly$\alpha$ profile. From the UVES data we obtain a very similar result ($\log N({\rm \ion{H}{1}})=20.44 \pm 0.05$, with the error dominated by systematic effects). {\color{black}However, it is difficult to be sure of the continuum over the damped Ly$\alpha$ profile in the echelle spectrum so we adopt the \citet{Pet94} and \citet{Pro01} value.}

That the singly ionized species do arise in a neutral region, so with the temperature
established above, in the WNM, is strongly indicated by 
the fact that it is the only component which is likely to give rise to the damped Ly$\alpha$ line.
Further, the corresponding \ion{Al}{3} lines are weak, and from Table \ref{tab:CAl} 
the \ion{Al}{3}/\ion{Al}{2} ratio is $-0.7$ dex, so the singly ionized species dominates. 
There is also no sign of a significant component of \ion{Si}{4} at the \ion{Si}{2} 
redshift, though it would fall between the two 
\ion{Si}{4} components given in the table. There is a \ion{C}{4} component at a redshift 
which, within the rather large errors, is consistent with the low ionization redshift. 
However the Doppler parameter is high so it is unlikely to be closely associated with
 the region of interest. 

If the ionization correction is negligible for the low ionization component, 
then the abundances relative to the solar photosphere
\citep{Asp09} are ${\rm [Fe/H]}=-2.59\pm 0.06$, ${\rm [Si/H]}=-2.29\pm 0.06$ and 
${\rm [O/H]}=-2.28\pm 0.08$.  Using the turbulent-plus-thermal
value for the Doppler parameters, ${\rm [C/H]}=-2.64\pm 0.06$, so ${\rm [C/O]}=-0.36$. 
These abundances and column densities are consistent with those given by  \citet{Pet08}, 
who assumed a common Doppler parameter of $b=6.5$\,{\kms} for all ions.
The \ion{Fe}{2} and \ion{Si}{2} values are also similar to those reported 
for this system by \citet{Pro97, Pro01}.

\subsection{Is the result reliable?}
\label{ss:systematics}

The temperature estimate here is {\color{black}12200K}, but the VPFIT error estimate of 3200K
is large, so at the $1\sigma$ level 
${\color{black}9\times 10^3<T<1.54\times 10^4}$K. When the errors are roughly comparable with 
the best estimate value one might be concerned at the reliability of 
the error estimate, since e.g. negative temperatures are unphysical. {\color{black}This is 
not quite the case here, but to further check we have generated artificial spectra using the 
best-fit parameters given in Table \ref{tab:sife}, convolved these spectra with the appropriate instrument profiles, added 
noise to mimic that in the original data, and determined the fit parameters from these spectra using the 
same fit regions and allowance for continuum and zero level adjustments as we did for the original data. 
From 2000 trials we find that the mean fitted temperature is $12270$\,K, with a standard deviation 
of $3350\pm 55$\,K. The expected error in the mean temperature estimated from the 2000 trials is then $3350/\sqrt{1999}=75$\,K. This mean temperature is therefore not significantly different from the input value, while the standard deviation
is a little higher than the mean VPFIT estimate of 3200\,K.}

\begin{figure}
\includegraphics[width=85mm,angle=0]{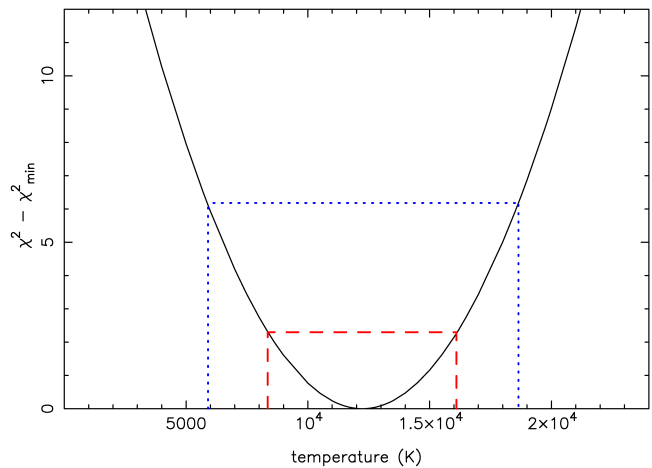}
\caption{Voigt profile fit $\chi^2$ values as a function of temperature from fits to
\ion{O}{1}, \ion{Si}{2} \& \ion{Fe}{2} lines. {\color{black} The dashed 
lines show the $\Delta\chi^2$ level for the $1\sigma$ range, and the dotted lines
for $2\sigma$, for two degrees of freedom.}
The $\chi^2_{\rm min}$ value is {\color{black} 893.39 (811 d.f.)}.}
\label{fig:chisq}
\end{figure}

{\color{black}Another way of checking the results is to derive}
the best fit $\chi^2$ values for a range of assumed temperatures $T$, and then use the values of 
$\Delta\chi^2(T)=\chi^2(T)-\chi^2_{\rm min}$ to estimate the error ranges. 
The method is described by \citet{Lam76}. In the application here we have 
at least two closely correlated (or anti-correlated) variables, the temperature and 
the turbulent Doppler parameter, so for a given probability $P$ we look for the temperature range
for which $\Delta\chi^2(T)$ is less than the value of $\chi^2$ corresponding to  
$P$ for two degrees of freedom. Under this assumption, for a degree of confidence 
of 68.3\% ($1\sigma$ for a normal distribution) 
$\Delta\chi^2(T)<2.3$, and for a $<2\sigma$ deviation for a two-sided 
threshold) $\Delta\chi^2(T)<6.2$. The results are shown in
Fig.~\ref{fig:chisq}, from which it is evident that the best fit value for the temperature
is in good agreement with that obtained from VPFIT, {\color{black}but the $1\sigma$  
error estimates from this method, $\pm 3900$\,K, are a little pessimistic, with those from this procedure being 
a little over $\sim 15\%$ larger than those from the simulations.}

These analyses preclude the possibility that a significant amount of 
the absorption arises in cold gas. If we take {\color{black}1000\,K as a somewhat generous} 
upper limit for 
CNM gas, then using the Doppler parameters are almost but not quite the same 
for the three ions, we find $\Delta\chi^2 = {\color{black}18.86}$. The formal probability of 
this occurring by chance is ${\color{black}1.4\times 10^{-5}}$.

{\color{black}Another way of establishing the probability that the material is 
WNM rather than CNM
is to generate synthetic spectra with the best-fit parameters for an assumed temperature
of 1000\,K, which we take as the upper limit for the CNM. These were $\log N($\ion{O}{1}$)=15.23$ (cm$^{-2}$)
with $b($\ion{O}{1}$)= 5.99$\,{\kms}, $\log N($\ion{Si}{2}$)=13.69$, $b($\ion{Si}{2}$)=5.95$\,{\kms} and
$\log N($\ion{Fe}{2}$)=13.36$, $b($\ion{Si}{2}$)=5.93$\,{\kms}. As in the simulations described above, we convolved these synthetic spectra with the appropriate instrument profiles, added noise to mimic 
that in the original data, and determined the fit parameters from these spectra using the same 
fit regions and allowance for continuum and zero level adjustments as we did for the original data. We also did 
this for 2000 trials, and find in only one case was the fitted temperature above 10000\,K, with none above 11000\,K. Since our temperature estimate was 12200\,K, we conclude that the probability that we are observing CNM gas is $< 0.0005$.}

Since we have chosen lines which sample different parts of the curves of 
growth we do not expect that our results will depend strongly on the chosen 
instrument profile.  However, since the absorption lines are resolved there could be some 
dependence on its assumed shape and width. We have undertaken some tests and 
find that this is not an important consideration. These were:

\begin{enumerate}



\item As a first check we note that for lines in the 
rest wavelength range 1140 - 1350\,{\AA} and for \ion{Si}{2} 1808 the seeing limited 
($\sim 0.5$ arcsec) and slit limited 
($\sim 1$ arcsec) exposure times were comparable, in the range 1.2:1 to 0.7:1. 
For \ion{Fe}{2} 1608, the \ion{Fe}{2} lines above 2300\,{\AA}, and \ion{O}{1} 1039 
the ratio is $\sim 0.5$:1 or less.
Since the inferred Doppler parameter for \ion{Fe}{2} is the lowest and most 
of the lines used in its determination have the broadest instrument profiles, 
it suggests that the raw instrument 
profile is not propagating directly into the inferred line widths. 

\item We have further checked that the temperature estimate is 
largely independent of the assumed instrument profile by performing 
the Voigt profile fits with different weights
for the 0.5:1 arcsec components. We also used single Gaussian point-spread functions 
for the whole spectrum, with widths ranging 
from 4.0 - 8.0 {\kms} FWHM. While the details changed, and at the ends of 
the Gaussian ranges the fits were not really acceptable, the temperature 
estimate proved robust with
$11000 < T < 13000$\,K. The error estimates were similar in all cases, 
at $\simlt 3600$\,K.

\end{enumerate}

{\color{black} While the narrow lines in the Ly$\alpha$ forest are easily separable from the
broad Lyman lines, the effective continuum for these narrow lines is affected by the
Ly$\alpha$ absorption. The fitting procedure does take account of the additional uncertainties
given the fit model, but, since the \ion{O}{1} 1039 line is strongly blended with Ly$\alpha$, as a rough consistency check we have refitted using only \ion{Si}{2} and
\ion{Fe}{2} to estimate the temperature. Not surprisingly, since the total ion mass range is lower by almost a factor of two, the error estimates increased 
significantly. We obtained $b_{\rm turb}=5.2 \pm 0.3$\,\kms and $T=1.7\pm 0.7\times 10^4$\,K.
So the best fit is at a higher temperature, but without the \ion{O}{1} the significance level 
for the material to be in the WNM rather than the CNM falls to $2.3\sigma$.}

\begin{figure}
\includegraphics[width=85mm,angle=0]{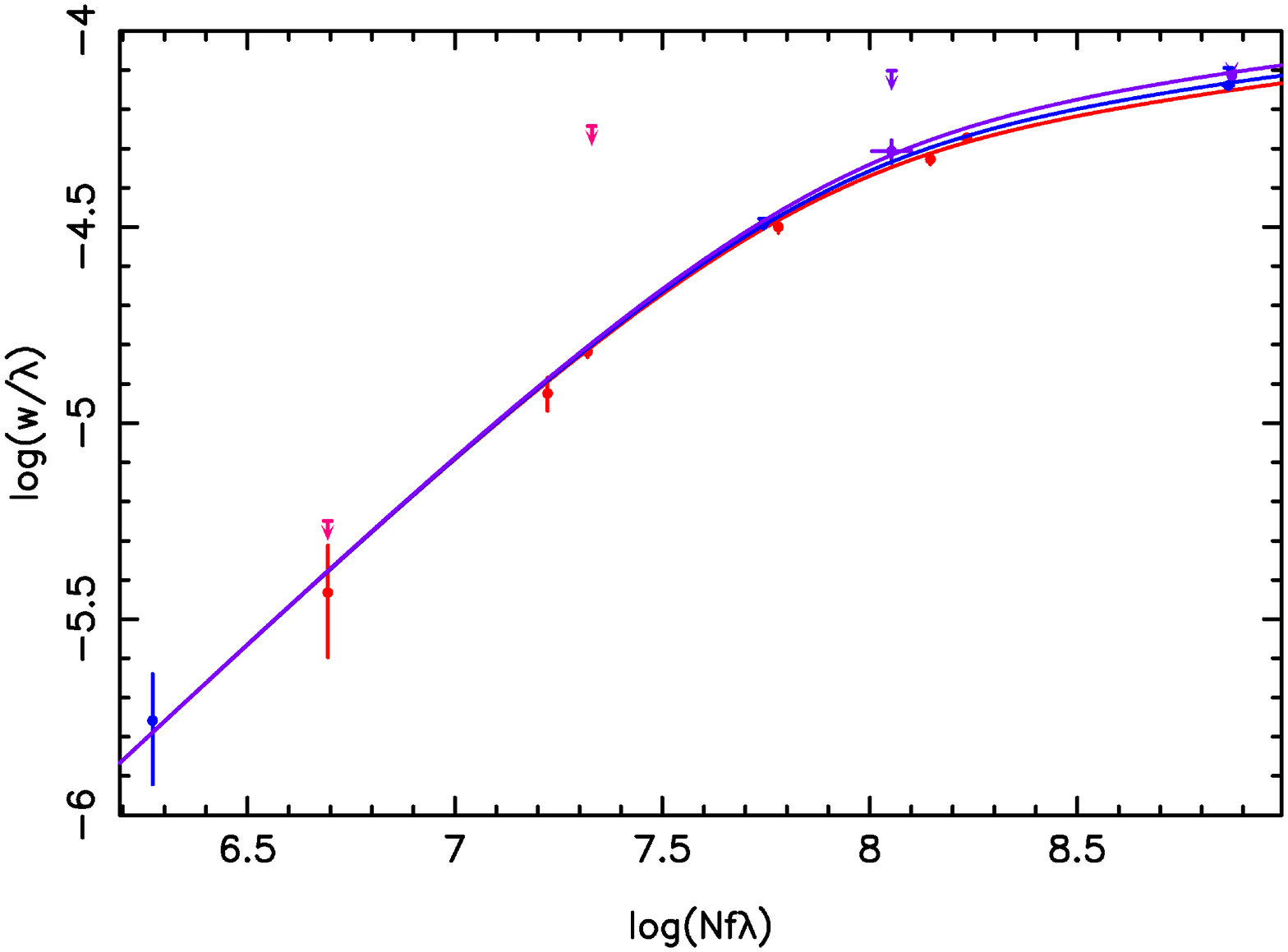}
\includegraphics[width=85mm,angle=0]{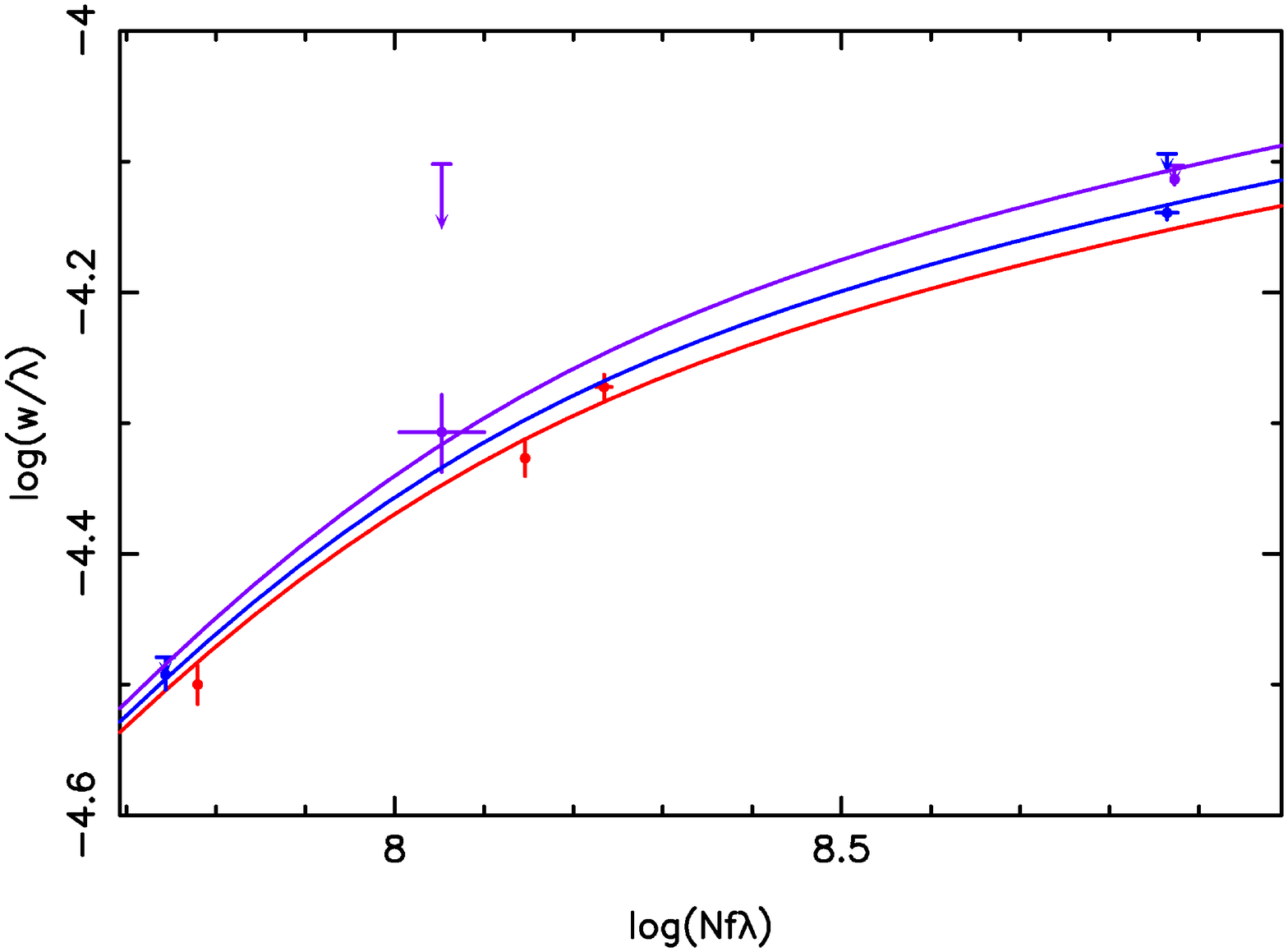}
\caption{Curves of growth {\color{black} using the fitted Doppler parameters given in Table \ref{tab:sife} for}
\ion{O}{1} (purple), \ion{Si}{2} (blue) and \ion{Fe}{2} (red), showing the estimated equivalent 
widths for the lines used. Upper limits are shown for blended lines, along with best estimates
(with errors) after the blends were removed. The horizontal error bars are based on the 
column density uncertainties from Voigt profile fitting. The lower panel gives detail near the
logarithmic parts of the curves of growth. See text for details. The units for $Nf\lambda$ are cm$^{-1}$.}
\label{fig:ew9}
\end{figure}

Ultimately, the most compelling evidence for a thermal component 
could come from curve-of-growth considerations, since then there are no uncertainties
involving the instrument profile. {\color{black}If} there is a thermal broadening component, 
then different mass ions which have observed transitions 
with a good range of $f\lambda$ values will lie on different curves of growth 
to fit the observed equivalent widths. The compatibility of VPFIT and the curve 
of growth analyses has been illustrated by \citet{Car11}. However they are not likely to be
completely equivalent. Changing from a description which for a set of lines assigns a single redshift, two parameters to describe the Doppler widths, and individual column densities for each ion to one where separate single parameters (the equivalent width) are derived for each 
line and then combining them as curves of growth will not necessarily yield exactly the same 
results. Also, blends with other lines are involved as in the Ly$\alpha$ forest, the individual equivalent widths are more difficult to determine. 

We have estimated the equivalent widths for the lines used by adding the contributions
from all pixels wholly within $\pm 20$\,{\kms} of the line centres, with error estimates obtained 
by summing the pixel contribution errors in quadrature. In this case possible blending 
is particularly important, since only lines with rest wavelengths
$> 1410$\,{\AA} are clear of the Ly$\alpha$ forest, and so both \ion{O}{1} lines are possibly blended and for \ion{Si}{2} only the line at 1808\,{\AA} is clear. \ion{Fe}{2} has several unblended
features with a range of $f\lambda$ values, but the transitions at 1260\,{\AA} and shortward are blended. So for the curves of growth based on the Doppler parameters given in Table 
\ref{tab:sife}, shown in Fig. \ref{fig:ew9} several of the lines appear as upper limits where
they are blended. For these we have also estimated equivalent widths over the $\pm 20$\,{\kms}
range after removing 
extraneous blends using profiles generated from the values given in that table. These are shown
as points near the curves underneath the limit markers. These curves of growth do not
cover all the lines used in the Voigt profile fits. \ion{Fe}{2} 1260 is close to \ion{Si}{2} 1260, and so was omitted here. Also \ion{Fe}{2} 1144 is close to a strong absorption feature, so while its upper limit is shown (at $\log Nf\lambda=7.33$) its corrected equivalent width was subject to large
errors so it was omitted. The point at $\log Nf\lambda=7.32$ near the curve corresponds to 
\ion{Fe}{2} 2374, which is free from contaminating blends.

While Fig. \ref{fig:ew9} illustrates that the profile fitting results are consistent with a curve of growth analysis with different curves for different mass ions, the curves of growth alone do not
require different curves for the various ions. They allow both fully turbulent and the mixed turbulent-plus-thermal broadening solution given above. We suspect this broader tolerance range comes from the broader error ranges allowed by this independent line approach. {\color{black} This is improved a little by changing the pixel weighting by fitting independent Voigt profiles to each component and then estimating equivalent widths from the line parameters, but that did not reduce
the errors by very much. To obtain a significant result we require stronger constraints e.g. by having a common Doppler parameter and column density for all lines of a given ion, but then we lose the independent errors so the curve-of-growth picture is less useful.} Curves of growth based directly on the fitted ions were, not surprisingly, compatible only with some WNM thermal broadening component. 

{\color{black} Fig. \ref{fig:ew9} also illustrates another point which should be
highlighted. Where the lines are narrow, a reliable Doppler parameter determination for any ion requires a number of transitions with a range of $f\lambda$ values which cover different parts of the curve of growth. At the
least not all the transitions should be only on the logarithmic or the linear parts of the curve since in
either case the Doppler parameter is poorly constrained. Also, if ions with single transitions (or transitions with similar $f\lambda$ values) are available they should not generally be included in the joint turbulent/thermal fit. Since the column density for each is then a free parameter, they provide minimal additional
information on the intrinsic line widths. Where the lines are marginally resolved the results may even be
misleading if such lines are included. We illustrate this by removing \ion{O}{1},  for which two lines 
with $f\lambda$ values separated by 0.82 dex were used, from the temperature analysis (Table \ref{tab:sife}) and try using \ion{C}{2} in its stead. \ion{Si}{2} and \ion{Fe}{2} together do not constrain the temperature very well, so \ion{C}{2} could, in principle, provide some improvement.
However, since the two \ion{C}{2} lines at 1036 and 1334\AA have similar $f\lambda$ values the result is critically dependent on the assumed line spread function (LSF). The nominal result for our best estimate of the instrumental profile is $T=24000\pm 3500$\,K. If we choose an instrumental FWHM for both \ion{C}{2} lines of 5.5\,{\kms} then the best fit temperature is $T=2.44\pm 0.36\times 10^4$\,K, while a FWHM of 8.0\,{\kms} gives $T=9.6\pm 4.8\times 10^3$\,K (in both cases the error estimates are those given by VPFIT). So, unlike the case where \ion{O}{1} was used, the result obtained with \ion{C}{2} is dominated by uncertainties in the instrument profile.}

The results above are based on a single component model with a single temperature 
and Gaussian turbulent velocity distribution. Given the apparent differences
in line widths between the various ions of different atomic masses it is hard 
to see how any multicomponent model could give significantly different 
temperature estimates. We did attempt two- and three-component models 
constrained to have only turbulent broadening in each velocity component, 
and each was a significantly poorer fit to the data than the model with 
some thermal broadening. 

We also tried a two-component turbulent-plus-thermal 
model and, not surprisingly, one component still had significant 
thermal broadening while another was completely
turbulent. Because two close velocity components provided the best fit in this model there is considerable degeneracy between them, and so the error estimates in individual quantities are 
so large as to make them almost meaningless. 

A possibility is that some of the \ion{O}{1} absorption might arise in a 
fully neutral region which is distinct from the region giving rise to the other lines, and then any small
velocity shift between the two might be interpreted as a broadening of the \ion{O}{1} lines.
However, this is unlikely since there is no detectable redshift difference between
the \ion{O}{1} and other species, and nor is there any sign of \ion{C}{1} which would also
arise in such a neutral component.

{\color{black}\subsection{Other temperature indications}
\label{ss:STIScomp}
}

{\color{black} \citet{Pet01} describe an analysis to determine the D/H ratio in this
system using a spectrum obtained with STIS on the HST covering the 
high order Lyman lines 
from \ion{H}{1} 926 down beyond the Lyman limit. This spectrum covers the 
wavelength range 2775 - 2860 {\AA} with a continuum S/N$\sim 6$ per 9.5 {\kms} pixel. 
Since the Ly$\alpha$
line alone gives the \ion{H}{1} column density from the damping wings, and the high order
Lyman lines will fall on the logarithmic part of the curve of growth, it 
should be possible to 
estimate the \ion{H}{1} Doppler parameter assuming that the absorption 
arises in a single component.  With hydrogen the mass range is much greater than
is available with heavy elements alone, so despite the poor S/N and resolution it should further
constrain the temperature estimate. \citet{Pet01} used VPFIT on these data and estimated a temperature of $\sim 11000$\,K assuming a thermal component of 5 {\kms}. This is in extremely good agreement with the value we have derived, but we suspect part of this agreement is fortuitous.
Not only is the S/N poor in the STIS spectrum so blended Ly$\alpha$ at lower redshifts are difficult to identify, but also the continuum is very uncertain, the zero level not well determined 
(it is about 4\% too high in the Lyman limit region), and the line profiles not well sampled. 
From joint fits Including both the STIS and UVES data, with the assumed \ion{H}{1} column 
density $\log N({\rm\ion{H}{1}})=20.43$, we find that 
$5.2\simlt b_{\rm turb} \simlt 5.4$\,{\kms} and $9.6\times 10^3 \simlt T \simlt 13.6\times 10^3$\,K. Much of the range comes from solutions involving different zero level and blending assumptions. We have not explored these very far, so the parameter ranges should be taken as indicative rather than firm, but they suggest that we have gained some precision but  not  a large amount. The  deuterium column density 
$\log N({\rm\ion{D}{1}})$
lies in the range 15.2 - 15.9 (cm$^{-2}$), so the possible error range for D/H is larger (by a factor of almost 3) than suggested by \citet{Pet01}. This reinforces the \citet{Pet08} suggestion that the earlier error estimates had been too small. }

There is another approach which can yield temperature limits to DLA absorbers, so we 
have explored its applicability here. 
\citet{Leh08} have used the \ion{C}{2}$^*$/\ion{H}{1} column density ratio
in a component of a redshift $z\approx 2.4$ DLA and the models
of \citet{Wol03} to show that the neutral gas in that system must be warm.
The reason in that case is that the \ion{C}{2}$^*$ 1335 is so weak relative to 
\ion{H}{1} Ly$\alpha$ that \ion{C}{2}\,158$\mu$m cannot be a dominant coolant so, unless 
the star formation rate is low, the absorption must be in WNM gas. Since we have shown
that the gas in the case studied here is WNM the \ion{C}{2}$^*$/\ion{H}{1} is also likely to be low. 
The upper limit to the \ion{C}{2}$^*$ column density in this system is
$\log N$(\ion{C}{2}$^{*}$)$< 12.75$ ($2\sigma$, Table \ref{tab:CAl}) and with 
$\log N$(\ion{H}{1})$=20.44$ the \ion{C}{2} 158\,$\mu$m cooling rate is
$\log l_c < -27.2$ erg s$^{-1}$ (H atom)$^{-1}$. This limit is too high 
by about 0.4 dex to require a WNM, but is consistent with it.

\section{Implications}
\label{sec:tempimp}

{\color{black} Some previous temperature estimates at high redshift have been based on
\ion{C}{1} and provided upper limits of a few hundred Kelvin \citep{Jor09, Car11}, so 
indicative of CNM gas. WNM material was found by \citet{OMe01} using a long atomic mass baseline from \ion{H}{1} to effectively \ion{Si}{2}. Here we have shown that it is possible
to use a range of ions from \ion{O}{1} to \ion{Fe}{2} to differentiate between cold and warm 
neutral media. A requirement is that the absorption lines used for each ion have a range of $f\lambda$ values sufficient to constrain their curves of growth, and if this condition is satisfied for all ions used
then the results will not be strongly dependent on the instrumental point-spread-function. So in favourable cases it should be possible to obtain temperature estimates
without the large mass range provided by hydrogen and deuterium with the heavy elements.
However, the lines which are used should not be strongly blended, and the turbulent
component should not predominate. The $b\sim 5$\,{\kms} found in the case described
here is probably close to the upper limit for current echelle data. Suitable cases may still 
be quite rare, since most of those in the literature exhibit complex line profiles. A preliminary
examination of over 20 UVES spectra of DLAs indicates that for $\simgt 90$\% of them the components 
are too  badly blended to consider further.}

Much of the emphasis on high redshift absorption systems has been 
to examine the redshift evolution of the heavy element abundances, and most of these, 
e.g. by \citet{Pro03}, have relied
mainly on the apparent optical depth method (AODM) \citep{Sav91, Sem92}. 
This method generally uses transitions which are near the linear part of the curve of growth, 
and so the results are insensitive to the Doppler parameter or any velocity structure.
For DLA gas, abundant ions such as \ion{Si}{2} and \ion{Fe}{2} have 
transitions with a wide range of oscillator strengths so their column densities can be 
estimated reliably. For less abundant species such as \ion{Zn}{2}, \ion{Cr}{2} and
\ion{Ni}{2} the lines are almost invariably sufficiently weak that the method gives
satisfactory results. In principle the \ion{O}{1} column density could
be constrained by its lower oscillator-strength transitions, but these
fall in the Ly$\alpha$ forest, and so they are subject to 
greater uncertainties. 

{\color{black}In the case of the $z_{\rm abs}=2.076$  DLA in the Q$2206-199$ spectrum
the heavy element column densities are not sensitive to the line broadening model adopted since bulk motion is a significant
contributor to the line widths for all ions. However, this may not always be the case. For an example of how the assumed broadening function may affect the inferred column densities
we consider the case of \ion{C}{2} at $z=2.340$ towards J$0035-0918$ \citep{Coo11}. They 
found a Doppler parameter $b=2.36$\,{\kms} which they used for all ions. 
For the data described by \citet{Coo11}  only \ion{O}{1} and \ion{Si}{2} have transitions with a range of
$f\lambda$ values, and so these two species were used. From these, the material may be purely
turbulent, as assumed by \citet{Coo11}, purely thermal, or anything in between with a best fit temperature
of $T\sim 4000$\,K. If instead of a purely turbulent model we consider the other extreme of a purely thermal model (i.e. turbulent component
assumed to be zero) applied to those data, the temperature is $T=7.66\pm 0.57\times 10^3$\,K.  Using this thermal model the column densities of the heavier ions are close to the values given by Cooke et al., but now $\log N({\rm \ion{C}{2}})=14.47\pm 0.06$ (instead of $15.47\pm 0.15$ for a
CNM model) and $\log N({\rm \ion{O}{1}})=14.73\pm 0.06$ (CNM: $14.96\pm 0.08$). The very large difference in the \ion{C}{2} value arises because the 
\ion{C}{2} lines are saturated and on the logarithmic parts of the relevant curves of growth. Then [C/Fe]$=0.51\pm 0.10$ rather than 1.53, which is still marginally outside the nominal range $-0.1<{\rm [C/Fe]}<0.4$ found in other low abundance and very high redshift systems \citep{Pen10, Coo11b, Bec12}.
}

While the example above is an extreme one, it does reinforce the need for care in inferring
column densities, as opposed to lower limits as in e.g. \citet{Des01}, where only a few
saturated lines are available. 

\section{Conclusions}
\label{sec:concl}

Comparison of the line widths for a range of ions of different masses, \ion{O}{1}, \ion{Si}{2} and
\ion{Fe}{2}, in the $z_{\rm abs}=2.076$ 
system in Q$2206-199$ yields a temperature estimate of 
$12000\pm 3000$\,K. This is consistent with expectations of $\sim 8000$~K for the Warm
Neutral Medium in galaxies, but not with the material being predominantly in a Cold Neutral Medium. In this case there is a significant bulk motion common to all ions, 
which is normally modelled as turbulence. Independent of the nature of the bulk motions, it is difficult to avoid the conclusion that the intrinsic line widths increase as the atomic mass decreases,
and that there is a significant component for which temperatures are $\sim 10^4$\,K.

{\color{black} Previous temperature estimates have relied on the comparison of \ion{H}{1}
and \ion{D}{1} Doppler parameters with those from heavy elements \citep{OMe01,Pet01}. We have shown here that if the ions used cover an adequate mass range, e.g. \ion{O}{1} to \ion{Fe}{2},
and for each ion the transitions cover a wide enough range of oscillator strengths that different
parts of their curves of growth are adequately sampled, then useful temperature estimates may be obtained. If these conditions are satisfied, then the temperature will not generally depend 
strongly on the assumed line-spread-function (LSF) for the spectrograph. However, including
other ions where the conditions are not satisfied can lead to the result which depends strongly on the assumed LSF.}

A significant fraction of DLA gas may exhibit temperatures of
$\sim 10^{4}$\,K. This is predicted for gas with weak \ion{C}{2}$^{*}$ absorption
by two-phase models for DLAs \citep{Wol03}, and may
be indicated by the \ion{H}{1} 21 cm results \citep{Kan09}.
In such cases the inferred column densities for important species such as \ion{C}{2}, where
the two observable interstellar lines have similar oscillator strengths (or more strictly, $f\lambda$ values) can depend critically on the assumed broadening mechanism for narrow-lined systems.
For example, the very high carbon overabundance reported by \citet{Coo11} could be reduced by a factor of $\sim 10$, from $\sim 35$ to $\sim 3$, if the lines are purely thermally broadened. 

Finally, we restate something which is well-known but sometimes overlooked: the column 
densities for a given ion are uncertain and depend on the assumed Doppler parameter 
only when all the transitions for that ion are saturated and {\color{black} when
thermal broadening component could be significant}. In cases 
where the Doppler parameters give implausibly high temperatures then e.g. single curve of 
growth analyses such as that by \citet{Mor78} should give reliable results for the 
ion column densities. The column densities in the $z=2.076228$ absorber reported here
are also consistent with those reported earlier \citep{Pro97, Pro01, Pet08} because
there is significant bulk motion so a single Doppler parameter is a reasonably
good approximation. {\color{black} However if the Doppler parameters for the heavier metals are
$\simlt 3$\,{\kms} then thermal broadening may be important, and  should be allowed for in 
estimating column densites of abundant species such as \ion{C}{2} and \ion{O}{1}. It is 
unfortunate that \ion{C}{2}, a species of current interest in low metallicity cases,
provides only two ground-state transitions for interstellar medium absorption studies.
These are usually saturated and, if that is the case, any derived column densities are likely to be uncertain}. 

\section*{Acknowledgements}

We are grateful to Ryan Cooke for his very helpful comments and providing us with results from thermal
fits to his J$0035-0918$ data, to Xavier Prochaska for discussions, and also a referee
for some very useful comments. RFC is thanks the Leverhulme Trust for an emeritus 
fellowship for part 
of this investigation. GDB was supported by the Kavli Foundation. 
RAJ acknowledges 
support from  the UK Science \& Technology Facilities Council 
through a grant to the Institute of Astronomy, Cambridge, and in Hawaii the US National Science Foundation through their fellowship program. MTM thanks the Australian Research Council for a QEII
Research Fellowship (DP0877998).  AMW acknowledges support by the NSF through grant
AST-1109452.

\bsp

\label{lastpage}

\end{document}